\documentclass[%
 preprint,
 amsmath,amssymb,
 aps,
]{revtex4-1}

\usepackage{graphicx,graphics}
\usepackage{dcolumn,comment}
\usepackage{subfigure}
\usepackage{booktabs}
\usepackage{bm,mathtools}
\usepackage{hyperref}

\DeclarePairedDelimiterX{\norm}[1]{\lVert}{\rVert}{#1}

\begin{document}

\title{Predictors and Predictands of Linear Response in Spatially Extended Systems}

   \author{Umberto Maria Tomasini}\thanks{Email: \texttt{umberto.tomasini@epfl.ch}}
  \affiliation{Dipartimento di Fisica e Astronomia, Universit\'a di Padova, Padova, Italy}
  \affiliation{Institute of Physics, EPFL, Lausanne, Switzerland}
  \affiliation{Department of Mathematics and Statistics, University of Reading, Reading, UK}
  \affiliation{Centre for the Mathematics of Planet Earth, University of Reading, Reading, UK}
   
 \author{Valerio Lucarini}\thanks{Corresponding author. Email: \texttt{v.lucarini@reading.ac.uk}}

  \affiliation{Department of Mathematics and Statistics, University of Reading, Reading, UK}
  \affiliation{Centre for the Mathematics of Planet Earth, University of Reading, Reading, UK}
   
   
   \date{\today}




\begin{abstract}
    
The goal of response theory, in each of its many statistical mechanical formulations, is to predict the perturbed response of a system from the knowledge of the unperturbed state and of the applied perturbation. A new recent angle on the problem focuses on providing a method to perform predictions of the change in one observable of the system by using the change in a second observable as a surrogate for the actual forcing. Such a viewpoint tries to address the  very relevant problem of causal links within complex system when only incomplete information is available.  We present here a  method for quantifying and ranking the predictive ability of observables and use it to investigate the response of a paradigmatic spatially extended system, the Lorenz '96 model. We perturb locally the system and we then study to what extent a given local observable can predict the behaviour of a separate local observable. We show that this approach can reveal insights on the way a signal propagates inside the system. We also show that the procedure becomes more efficient if one considers multiple acting forcings and, correspondingly, multiple observables as predictors of the observable of interest. 
\end{abstract}

\maketitle

\section{Introduction}
\subsection*{Elements of Response Theory}
\emph{Response Theory} is an area of statistical physics that provides general methods for predicting the changes in the statistical properties of an observable of interest $\Psi$ from the knowledge of the applied perturbation and of the statistical properties of the unperturbed system.  
%
The expectation value of $\Psi$ in the perturbed system is expressed as a perturbative series where the zeroth-order term is the expectation value of $\Psi$ in the unperturbed system. The higher-order terms are expressed in terms of \emph{response functions} which contain information about the higher order statistics  of the unperturbed system and the applied forcing

A cornerstone in the development of response theory came with the work by Kubo \cite{k1,k2}, who considered the case of weakly perturbed systems near \emph{thermodynamic equilibrium}. When a system is in this kind of steady state, it obeys detailed balance and features no net currents. 
In  Kubo's theory, from the knowledge of the statistics of the unperturbed system, and in particular of the correlations describing its spontaneous fluctuations, it is possible to compute, in linear approximation, the response of the system to any  (weak) perturbation.  This is the key idea behind the fluctuation-dissipation theorem (FDT) \cite{k2}, which establishes a link between the forced and free fluctuations in the perturbative regime 
\footnote{Kubo's theory initially received a vehement criticism by Van Kampen \cite{vk}; see in \cite{lrt} a clear counter-argument to Van Karmen's argument.}. A generalised version of the FDT valid for higher moments has been proposed by \cite{Lucarini_2012} in the spirit of Zubarev's generalization of Kubo's results \cite{Zubarev_1974}. We will discuss from now on the special and very relevant case of linear response theory (LRT) and of its applications.



The analysis of the response to perturbations of systems that are far from thermodynamics equilibrium requires a more general approach than what provided by Kubo's theory. Indeed, nonequilibrium systems are ubiquitous and their investigations has both great theoretical relevance and allows one to study many important real-life examples in material science, physics, ecology, fluid dynamics, climate science, biology, among others. Often, one says that after transients have died out, in absence of time-dependent forcings, a nonequilibrium system is in the so-called nonequilibrium steady state, where detailed balance is not obeyed, and, in many cases, dissipative processes are associated with a contraction of the phase space and with the production of entropy \cite{Gallavotti2014}. Studying the response of NESS systems to perturbations is of great relevance both for theory and applications. At this regard, a rigorous and crucial development in the context of deterministic dynamics was provided by Ruelle \cite{r0,r1,r2,ruelle}, who rigorously derived a LRT for smooth observable of Axiom A dynamical systems. Ruelle's results have been been recast and extended by methods of functional analysis \cite{BL07,B08}. Roughly speaking, Axiom A systems provide an extremely useful example of chaos  characterized by the presence of a clear separation between the expanding and contracting directions in the tangent space, formalized by the concept of uniform hyperbolicity. Crucially, Axiom A systems possess a special ergodic invariant measure - the so-called Sinai-Ruelle-Bowen measure \cite{r0} - that makes them excellent candidates for describing non-equilibrium physical systems. While Axiom A systems are mathematically quite special, their practical relevance is clarified by the \emph{chaotic hypothesis} proposed by Gallavotti and Cohen \cite{g1,g2}, which can be see an extension of the ergodic hypothesis to the non-equilibrium case.

The chaotic hypothesis is supportive of the fact that LRT should \textit{de facto} work in a large class of chaotic dynamical systems. As opposed to the equilibrium case, general non-equilibrium dynamical systems obeying Axiom A dynamics possess an invariant measure that is singular with respect to Lebesgue and is supported on a strange attractor. Ruelle proved that the response operator is given by the sum of two contributions. One is related to the dynamics on the unstable and central manifolds and can be framed as a FDT result, while the second one is related to the dynamics occurring on the stable manifold, and cannot be framed as a FDT result because of the non-smoothness of the measure \cite{ruelle}. In other words, the natural fluctuations are not equivalent to the forced perturbations along the stable directions \cite {luc10,luc11,sarno}. The direct computation of these two contributions is far from trivial \cite{abram,abram1, baiesi}, but the use of adjoint and shadowing methods has recently led to promising results in this direction \cite{wangQ1,wangQ2,ni}.

LRT can also be rigorously established for  stochastic dynamical systems \cite{Hairer2010,h1,baiesi,Seifert_2010}. Under rather general hypotheses, adding noise makes the invariant measure absolutely continuous with respect to Lebesgue, so that the FDT fully holds \cite{cola}. In this setting, the obtained formula is called the Kubo-Agarwal formula, which reduces to the Kubo formula for equilibrium systems.  The addition of a noise term has to be justified by the nature of the considered problem. This stochastic perspective becomes relevant in many complex systems, where the focus is on coarse-grained dynamics, which is effectively stochastic as a result of the presence of microscopic degrees of freedom. Note that the coarse-grained dynamics is in general non-markovian, with memory effect becoming negligible in the limit of infinite time-scale separation between the fast and slow variables \cite{zwanzig_memory_1961,mori_transport_1965,wou,CLW15a,CLW15b}. 

The relationship between response theory in deterministic and stochastic systems has been thoroughly discussed in \cite{Wormell2019}, where the authors also propose a very well-developed justification, not based on the chaotic hypothesis, for the broad pragmatic applicability of LRT in a vast class of deterministic chaotic systems.

Nowadays, LRT has an important role in the investigations of a vast range of systems see e.g.  \cite{Ottinger2005,lrt,baiesi,Lucarini2017,Cessac2019,Sarracino2019,GhilLucarini2020,Lembo2020}. As an example, in the case of climate science, LRT makes it possible to perform climate change projections using climate  models of different levels of complexity. This amounts to computing the time-dependent measure supported on pullback attractor of the climate \cite{Chekroun2011} by constructing response operators for a suitably defined reference climatic state  \cite{GhilLucarini2020}. This viewpoint allows one to investigate ways to control the future pathways of climate change \cite{Aengenheyster2018} and to make an assessment of potential and pitfalls of geoengineering strategies \cite{Bodai2020geo}. 

Recently, LRT has been extended in such a way that explicit formulas are given for describing how adding a forcing to a system changes its $n-$point correlations \cite{Lucarini_2017}. Another recent application of LRT has focused on detecting and characterising phase transitions in a network of coupled identical agents undergoing a stochastic evolution \cite{Lucarini2020PRSA}.   A recently published special issue showcases several emerging areas of applications for LRT  \cite{Gottwald2020}. 

Majda and Qi have recently shown the existence of a coherent thread connecting LRT, sensitivity analysis, model reduction techniques, uncertainty quantification, and control of high-dimensional chaotic and stochastic systems \cite{Majda2018,Majda2019}. The theme of studying the response using incomplete information on the system is the main topic of the present contribution .

\subsection*{Predictors and Predictands}

A different angle on the problem of defining the response of a system to perturbations proposes a fairly general method that allows one to relate the response of different observables of a system undergoing a perturbation. This method could be useful in  situations where we need to interpret experimental data and we do not have necessarily perfect knowledge of the  properties of the system - e.g. we do not know the specific features of the applied forcing, have access to only a limited number of degrees of freedom of the system, or ignore altogether the evolution equation of the system itself -  but can measure instead  multiple observables at the same time \cite{luc0}. This is closely related to finding solutions to nonlinear rational least squares problems \cite{Gustavsen1999,Berljafa2017}.

The goal is to understand to what extent we can use perturbed observables as surrogates of the perturbation to reconstruct the time behaviour of other observables. It turns out that, if we know the time behaviour of one observable $\Psi_{1}$, we can always reconstruct diagnostically, within linear approximation,  the time behaviour of another observable $\Psi_{2}$ through a surrogate response function. 

Instead, if the goal is to actually predict the future state of the observable $\Psi_{2}$, by means of a prognostic relation, it turns out that \emph{not} all choices of predictor are equally successful to perform the prognosis of a given predictand, because the surrogate response function might be, as opposed to the standard response (Green) function, not causal, i.e. its support is not limited to non-negative times. 

The analysis performed on the Lorenz '96 model \cite{lorenz1,lorenz2,lorenz3} in \cite{luc0} explicitly showed that the response of an observable $\Psi_{1}$ to a given perturbations could be predicted by the response of a certain observable $\Psi_{2}$ but could not be predicted by the response of a third one $\Psi_{3}$, because in the latter case the surrogate response function has a non-causal component.

\subsection*{This Paper}
Following the seminal contribution by Granger \cite{Granger1969}, more and more attention has been recently paid to finding rigorous ways for defining and detecting causal links within complex systems \cite{Pearl2009}, and climate science has been a very successful fertile of application of such ideas \cite{Runge2014,Hannart2016,Saggioro2019,Runge2019}. 

In this paper, we address this class of problems using a different viewpoint. We want here to make a step forward compared to \cite{luc0} with the goal to better quantify the skill of different observables in predicting the response of a given observable. Here, we propose a way to evaluate \textit{how much} their corresponding surrogate response functions are not causal.  
This problem can emerge in a variety of situations where we have more non-predictive surrogate response functions and we want to choose between them the one which provides the best prediction. For example, we could have a set of observables $\{\Psi_{1},...,\Psi_{m}\}$ and we could aim at finding out which one(s) can be used to  predict most accurately the response of another observable $\Psi_{a}$. Any observable whose surrogate response function is predictive is equally good at this regard. If only one among the $m$ observables features a predictive surrogate response function, the choice is obvious. The choice becomes less straightforward when all the observables $\{\Psi_{1},...,\Psi_{m}\}$ are not predictive. We would like to have a quantitative method that allows us to choose the most predictive one among them. Another setting where such a method could be helpful is when we want to 
understand whether an observable $\Psi_{a}$  predicts better $\Psi_{b}$ or viceversa, as a better predictive power can be linked to a causal link or a flow of information with a definite verse from an observable to the other. 

Specifically, we investigate causal links in the context of a spatially extended system undergoing chaotic dynamics, namely, the  paradigmatic Lorenz 96 (L96) model \cite{lorenz1,lorenz2,lorenz3}. We investigate the response of the system to localised forcings, and we observe the system by looking at its local properties in different regions. We assess whether the response of local observables  can be used to predict the response of other local observables of interest. One of the goals of this analysis is to see whether we can detect a clear indication of the way signals propagate inside the system. We also explore the possibility suggested in \cite{luc0} that the predictive skill of the surrogate approach proposed here could dramatically improve if one perturbs the system with multiple forcings and uses multiple observable as surrogate predictors. 


The rest of the paper is structured as follows. In {Section \ref{srt}}, after briefly reviewing the surrogate LRT \cite{luc0}, we present the predictability index (PI), which  aims at quantifying how much a surrogate response function is non-predictive. 
We will remark that the presence of such an index provides the opportunity to build an \emph{hierarchy} of observables in terms of their predictive power of other observables. In {Section \ref{l96}} we apply the surrogate LRT on the L96 model, considering local perturbations and local observables. We will also explore the impact of adding information gathered from global observables to predict the local forced response. In  {Section \ref{conclusions}} we summarise the main findings of this study and present perspectives for future research. A set of Appendices provides supplementary material of possible interest for the reader. In Appendix \ref{exampleR} we provide an illustrative example to better explain the meaning of the PI. 
In  Appendix \ref{linear} we provide evidence of the fact that in our experiments and data analysis we are in a linear regime of response. 
In Appendix \ref{asymp} we clarify some asymptotic properties of the response of the Lorenz '96 model, while in Appendix \ref{sing_one} some specific properties of the surrogate response functions are discussed.

\section{Surrogate Response Theory} \label{srt}
Following Ruelle \cite{r1,r2,ruelle}, we recapitulate very informally some basic elements of LRT by studying the effect of perturbing an  Axiom A system of the form $\dot{\vec{x}}=\vec{F}(x)$, where $x\in \mathcal{M}$, a smooth compact manifold of dimension $D$. We introduce the following complex pattern of forcing, consisting in $N$ independent perturbations:
\begin{equation}
    F(x)\Rightarrow F(x) +\sum_{l=1}^{N}e_{l}(t)G_{l}(x)
\end{equation}
where $G_{1}(x),\ldots,G_{N}(x)$ are $D-$dimensional smooth vector fields while $e_{1}(t),\ldots e_N(t)$ are time patterns. In linear approximation, which is relevant in the case the applied forcings are small, it is possible to write the change in the expectation value of any smooth observable $\Psi$ as follows: 
\begin{equation}
\begin{aligned}
    \delta \langle \Psi \rangle (t) = \sum_{j=1}^N\int_{-\infty}^{\infty}d\tau e_j(\tau) \Gamma_{\Psi,G_j}(t-\tau),
    \end{aligned}
    \label{u155}
\end{equation}
where we have introduced the linear (Green) response functions $\Gamma_{\Psi,G_j}$, which mediate the effect of the time pattern of the perturbation $e_j$ at time $\tau<t$ on the observable $\Psi$ at time $t$. The response function can be seen as the expectation value in the unperturbed system of a complex observable that  depends on the applied forcing and on $\Psi$: 
\begin{equation}
    \Gamma_{\Psi,G_j}(t)= \Theta(t) \int \rho_0(dy)  G_j(y) \nabla_{y}(\Psi(y(t))),
    \label{u15}
\end{equation}
where $\Theta$ is the Heaviside distribution, which determines the causality of the response functions, $\rho_0$ is the invariant measure of the unperturbed system and $\Psi(y(t))=S^t\Psi(y)=\exp(\mathcal{L}_0 t) \Psi(y)$ is the value of the observable $\Psi$ at time $t$ following the evolution in time according to the dynamics of the unperturbed system, with initial condition $y$. We have that $S^t$ is the (unperturbed) Koopman operator and $\mathcal{L}_0=F\cdot\nabla$ is its generator. 
By applying the Fourier transform to Eq. \ref{u155}, we obtain the following identitys:
\begin{equation}
\begin{aligned}
    \delta \langle \Psi \rangle (\omega) = \sum_{j=1}^N e_j(\omega) \chi_{\Psi,G_j}(\omega),
    \end{aligned}
    \label{u155f}
\end{equation}
where $\chi_{\Psi,G_j}(\omega)$, $j=1,\ldots,N$ are the so-called susceptibilities. Since the response functions $\Gamma_{\Psi,G_j}(t)$, $j=1,\ldots,N$ are causal, under standard conditions of integrability the corresponding susceptibilities are analytic in the upper complex $\omega-$plane \cite{Nussenzweig1972,vbook}.

\subsection{Surrogate response functions}
A different angle of the problem in LRT has been introduced in \cite{luc0} where response relations \emph{between perturbed observables} are built. These relations can be useful in a large variety of contexts where the knowledge of the forcing is just partial, and we want to use perturbed observables to diagnose the state of other perturbed observables. We consider $N+1$ independent observables $\Psi_{1}(x),\ldots,\Psi_N(x)$.  In \cite{luc0} it is shown that it is possible to express the linear change of the expectation value of an observable $\Psi_{N+1}$ as a function of the ones of the other $N$ observables;
\begin{equation}
    \delta\langle\Psi_{N+1}\rangle(\omega)=\sum_{l=1}^{N}J_{N+1,l}(\omega) \delta\langle\Psi_{l}\rangle(\omega),
    \label{s19}
\end{equation}
where we take the surrogate of the $N$ forcings using the other $N$ observables through the surrogate susceptibilities $J_{N+1,l}$, $l=1,\ldots,N$. In \cite{luc0} explicit expressions for the surrogate susceptibilities $J_{N+1,l}(\omega)$ are derived:
\begin{equation}
    \left(\begin{array}{c}
	J_{N+1,1}(\omega) \\
	\vdots\\
	J_{N+1,N}(\omega)
\end{array} \right)=
\left(\begin{array}{ccc}
	\chi_{\Psi_{1},G_{1}}(\omega) &\ldots& \chi_{\Psi_{N},G_{1}}(\omega) \\
	\vdots&\ddots&\vdots\\
	\chi_{\Psi_{1},G_{N}}(\omega) &\ldots& \chi_{\Psi_{N},G_{N}}(\omega)
\end{array} \right)^{-1} \left(\begin{array}{c}
	\chi_{\Psi_{N+1},G_{1}}(\omega)\\
	\vdots\\
	\chi_{\Psi_{N+1},G_{N}}(\omega)
\end{array} \right).
\label{n202}
\end{equation}
Once we have obtained the surrogate response functions, we plug them into Eq. \ref{s19}, obtaining the following expression:
\begin{equation}
    \delta\langle\Psi_{N+1}\rangle(t)=\sum_{l=1}^{N}\int_{-\infty}^{+\infty}\,d\tau H_{N+1,l}(t-\tau) \delta\langle\Psi_{l}\rangle(\tau).
    \label{s20}
\end{equation}
where $H_{N+1,l}(t)$ is the inverse Fourier transform of $J_{N+1,l}(\omega)$, $l=1,\ldots,N$. 

Note that the previous relation does not involve the time patterns of the acting forcings. This has important practical consequences. If we are able to derive the functions $H_{N+1,l}(t)$ or $J_{N+1,l}(\omega)$, $l=1,\ldots,N$ from a probe experiment performed with a (broadband) time pattern of forcing, we can use Eq. \ref{s20} to reconstruct the time evolution of $\delta\langle\Psi_{N+1}\rangle(t)$ for any time pattern of the forcing using as input the time evolution of $\delta\langle\Psi_{j}\rangle(t)$, $l=1,\ldots,N$. This will investigated in the next Sect. \ref{num}. This inverse problem can also be approached by considering monochromatic forcings and scanning across frequencies, in the spirit of the algorithms proposed in \cite{Gustavsen1999,Berljafa2017}.

Equation \ref{s20} can be employed to diagnose, at the linear level in the perturbation, the time behaviour of $\Psi_{N+1}$ by means of other $N$ observables. On the other hand, it is not always possible to perform the prognosis of $\Psi_{N+1}$ using the same observables, which can be useful in prediction problems. This requires that the response functions  $H_{N+1,l}$ $\forall$ $l=1,\ldots,N$ are causal, i.e. their support is in the non-negative domain. 

Let's expand more on the case $N=1$, to better clarify some issues associated with the surrogate LRT. Equations \ref{n202}-\ref{s20} become:
\begin{equation}
    J_{2,1}(\omega)\equiv \frac{\chi_{\Psi_{1},G}(\omega)}{\chi_{\Psi_{2},G}(\omega)},\qquad \delta\langle \Psi_{2}\rangle(t)=\int_{-\infty}^{\infty}H_{2,1}(t-\tau)\delta\langle \Psi_{1}\rangle(\tau).
    \label{s3}
\end{equation}
The surrogate response function $H_{2,1}$ is predictive - i.e., it has support only on the non-negative domaion - if and only if its Fourier transform $J_{2,1}$ has no poles in the upper complex $\omega-$plane. Since the numerator $\chi_{\Psi_{1},G}(\omega)$ is analytic in the upper complex $\omega$ plane, loss of predictability is realised only if the response function $\chi_{\Psi_{2},G}(\omega)$ at the denominator has complex zeros in the upper complex $\omega$ plane, see discussion in \cite{luc0}. 

Indeed, we expect that for a given forcing not all choices of predictors and predictands are equally successful in terms of predictive power. For instance, if there is a causal relation in a feedback or a flow of information linking observable $\Psi_1$ and $\Psi_2$, one expects the presence of an asymmetry in the mutual predictive power. 


Lastly, we remark that the surrogate response function could have a singular component in 0, as it is noted in \cite{luc0}. There is a close link between the short-time behaviour of $\Gamma_{\Psi_{1},G}$ and the high-frequency behaviour of its Fourier transform:
\begin{equation}
 \Gamma_{\Psi_{1},G}(t)\approx \alpha_{\Psi_{1},G}\Theta(t)t^{\alpha} 	\Leftrightarrow \chi_{\Psi_{1},G}(\omega)\approx (\alpha_{\Psi_{1},G}\, \alpha!\, i^{\alpha+1})\frac{1}{\omega^{\alpha+1}} 
 \label{e100}
\end{equation}
As a consequence, it is possible to obtain the asymptotic behaviour of the Fourier transform of the surrogate response function $J_{2,1}$ using the asymptotic behaviour of the response functions. In particular, if for large values of $\omega$ we have that $\chi_{\Psi_{1},G}\approx1/\omega^{\alpha_{1}+1}$ and $\chi_{\Psi_{2},G}\approx1/\omega^{\alpha_{2}+1}$, we derive:
 \begin{equation}
 \begin{aligned}
  J_{2,1}(\omega)\underset{\omega\rightarrow \infty }{\propto}& \frac{1/\omega^{\alpha_{1}+1}}{1/\omega^{\alpha_{2}+1}}\\
  \underset{\omega\rightarrow \infty }{\propto} &\omega^{\alpha_{2}-\alpha_{1}}.
 \end{aligned}
 \label{n20}
 \end{equation}
If $\alpha_{1}<\alpha_{2}$  $J_{2,1}(\omega)$ diverges for large values of $\omega$, while it converges to a non-vanishing constant for $\alpha_{1}=\alpha_{2}$. Hence, in these cases the surrogate response function $H_{2,1}(t)$ will have a singular component $S_{2,1}(t)$ in $t=0$ because the Fourier transform of $(-i\omega)^{j}$ is $\delta^{j}(t)$, i.e. the $j-$th derivative of the delta function $\delta(t)$. We can then write:
\begin{equation}
    H_{2,1}(t)=S_{2,1}(t)+K_{2,1}(t),
    \label{w1}
\end{equation}
where $K_{2,1}(t)$ ($S_{2,1}(t)$) is the non-singular (singular) component. 
On the contrary, if $\alpha_{1}>\alpha_{2}$ the surrogate response function $H_{2,1}(t)$ has no singular component $S_{2,1}$. 

Note that the exponent $\alpha$ describing the short time behaviour of the response function controls how long it takes for the observable to feel the effect of the forcing. The higher the exponent, the slower is the build-up of the effect of the forcing on the observable. Hence, it makes sense to use $\Psi_{1}$ to predict $\Psi_{2}$ only if $\alpha_{1}\le \alpha_{2}$, i.e. if $\Psi_{1}$ feels the applied perturbation before or approximately at the same time as $\Psi_{2}$. From now on we then exclude the case $\alpha_{1}> \alpha_{2}$.

The part of the surrogate response function $H_{2,1}$ that is practically usable for predictions has support restricted to the non-negative domain and can then be expressed as follows:
\begin{equation}
    H_{2,1}^{c}(t)=\Theta(t)K_{2,1}(t)+S_{2,1}(t).
    \label{0f}
\end{equation}

In practice, since the outputs of numerical simulations have finite precision, $H^{c}_{2,1}(t)$ can be reconstructed from data as follows:
\begin{equation}
    H^{c}_{2,1}(t)\equiv \lim_{\varepsilon\rightarrow 0^{+}}\left (\Theta(t+\varepsilon)H_{2,1}(t) \right),
    \label{f0}
\end{equation}
where the $\varepsilon-$ regularization has been introduced to include in the definition of $H^{c}_{2,1}$ possible singular components of $H_{2,1}(t)$ at $t=0$.

\subsection{Quantifying the Ability to Predict} \label{ratio_method}

The presence of the non-causal component in the surrogate response function $H_{2,1}$ hinders the prediction of $\Psi_{2}$ at time $t$ using just the time behaviour of $\Psi_{1}$ up to time $t$. An interesting problem is to actually quantify the non-causal component of the surrogate response function. This quantification would allow to build an hierarchy of observables in terms of their predictive power of other observables.

From the discussion above, we then have that the surrogate response functions are of the  form:
\begin{equation}
\begin{aligned}
    H_{2,1}(t)=&S_{2,1}(t)+K_{2,1}(t)\\
    =&s_{2,1}\delta(t)+K_{2,1}(t),
\end{aligned}
    \label{w11}
\end{equation}
where the constant $s_{2,1}\in \mathbb{R}$ can also be zero. We propose to quantify the ability of the observable 1 to predict the observable 2 with the predictability index (PI), which is defined as follows:
\begin{equation}
    R(H_{2,1})\equiv\frac{\norm{K^{nc}_{2,1}(t)}_{1}}{\norm{ K^{c}_{2,1}(t)}_{1}+\norm{s_{2,1}(t)}_1},
    \label{t6}
\end{equation}
where:
\begin{equation}
    K^{c}_{2,1}(t)=\Theta(t)K_{2,1}(t),\qquad K^{nc}_{2,1}(t)=\Theta(-t)K_{2,1}(t),
\end{equation}
hence $K_{2,1}^{c}$ is the causal part of the non-singular component of the surrogate response function, while $K_{2,1}^{nc}$ is its non-causal part. Additionally, $\norm{\bullet}_{1}$ is the standard $L_1$ norm. The PI depends on the system under investigation, the space pattern of the forcing $G(x)$ and on the observables $\Psi_{1}$ and $\Psi_{2}$. 

The PI is non-negative and vanishes if and only if the surrogate response function is predictive, because its support includes only the non-negative domain. A large value for the PI indicates that the response of the observable 1 is a poor predictor of the response of the observable 2. Moreover, since this method revolves around the surrogate response function, it does not depend on the chosen time pattern. We will actually see the effectiveness of this indicator in the L96 model in Section \ref{l96}. A few pedagogical examples can also be found in Appendix \ref{exampleR}. 

We can generalize the indicator given in Eq. \ref{t6} to the case when we use $\Psi_l$, $l=1,\ldots,N$ observables as  predictors of the observable $\Psi_{N+1}$, as in Eq. \ref{s19}. For each surrogate response function $H_{N+1,l}(t)$, with $l\in\{1,...,N\}$, we define its singular part $S_{N+1,l}(t)$ and its non-singular part $K_{N+1,l}(t)$, which, in turn, can be split into the non-causal component $K_{N+1,l}^{nc}$ and the causal component $K_{N+1,l}^{c}$. We assume that all these surrogate response functions are of the type Eq. \ref{w11}. We then define: 
\begin{equation}
    R(\{H_{N+1,l}\}_{l=1,...,N})\equiv\frac{\sum_{l=1}^{N}\norm{K^{nc}_{N+1,l}(t)}_{1}}{\sum_{l=1}^{N}\left(\norm{ K^{c}_{N+1,l}(t)}_{1}+\norm{S_{N+1,l}(t)}_1\right)}.
    \label{t66}
\end{equation}

\section{The Lorenz '96 model}\label{l96}

\subsection{Model Formulation}\label{ll96}
The L96 model \cite{lorenz1,lorenz2,lorenz3} is a paradigmatic  model that provides a metaphor of some essential properties of the dynamics of the atmosphere. It is defined on a lattice of $N$ grid points and the evolution of the variables $x_i$, $i=1,\ldots,N$ is given by the following system of ordinary differential equations: 
\begin{equation}
    \dot{x}_{i}=x_{i-1}(x_{i+1}-x_{i-2})-\gamma x_{i}+F,
    \label{l1}
\end{equation}
where $F$ is a constant controlling the external forcing, $\gamma$ (usually set to a unitary value, as done also here) modulates the strength of the dissipation, and the quadratic term on the right hand side describes a non-standard advection process. The system obeys periodic boundary conditions, so that $x_{i-N}=x_{i}=x_{i+N}$ for all values of $i$. 
The qualitative properties of the dynamics of the L96 model changes dramatically accordingly to the values of $F$ and $N$. In particular, it can be shown that, when $\gamma=1$, the dynamics of the model is chaotic for $F\ge 5$ and the system becomes to a good approximation extensive as $N\ge 20$ \cite{galla_luc}. 
One can show that  travelling waves appear on top of the turbulent background in the chaotic regime, with phase velocity directed towards decreasing values of $i$ \cite{lorenz1}. The travelling waves are roughly preserved even in the chaotic regime. Instead, the group velocity, which marks the direction of the propagation of information within the system, is directed in the opposite direction, towards increasing values $i$ \cite{vissio2020}. Detailed analyses of the properties of the L96 model can be found in \cite{galla_luc,sterk,kekem}, where the reader can find also an extensive bibliography. Recently, two extensions of the L96 model have been proposed, one able to accommodate for a complex interplay between  dynamics and thermodynamics \cite{vissio2020}, and one featuring multiple competing attractors \cite{Gelbrecht2020}.


\subsection{Numerical Simulations} \label{num}
\subsubsection{Linear response functions}
We will now apply the formalism of the surrogate LRT to the L96 dynamical system Eq. \ref{l1}. We will set ourselves in the chaotic regime by choosing $N=36$ and $F=8$. In \cite{luc0}, the focus was on global perturbations, which impact directly all the variables $x_{i}$, $i=1,\ldots N$, and on global observables. We want now to take a different route, focusing on local perturbations and local observables, which had initially been proposed in \cite{sarno}. In particular, we  choose the following spatial pattern of applied forcing:
\begin{equation}
G_{i}(x)=\epsilon\delta_{ik}, \quad i=1,\ldots N
\label{l7}
\end{equation}
where $\delta_{ik}$ is the  Kronecker delta, which has unitary value if the two indices are identical and vanishes otherwise, and $\epsilon$ is a real number which measures the magnitude of the perturbation. Equation \ref{l7} implies that we apply an extra forcing only at  $k^{th}$ grid point. 
We will consider as observables of interest the dynamical variables $x_{j}$. With these choices of the perturbation and the observables, the problem we are addressing amounts to asking to what extent a perturbed variable at location $i$ can predict the future state of another perturbed variable at location $j$ after the system has been perturbed locally at location $k$. Given the discrete symmetry of the system, we expect that these properties will depend only on the relative position of $i$, $j$, and $k$, but, since the propagation of the information in this system is directional, we expect that they will not depend only on the distance between these locations.

Let $\Gamma_{i,k}$ be the response function of the perturbed variable $x_{i}$ to the perturbation with spatial pattern Eq. \ref{l7}, located in $x_{k}$. Along the lines of  \cite{luc0}, we estimate $\Gamma_{i,k}$ by considering a probe with time pattern a Dirac's delta: $e(t)=\delta(t)$. We then run a long simulation with a random initial condition using adaptive Runge-Kutta method of order 4 implemented through the Python function \texttt{solve\_ivp}. We discard an initial transient of length $T_{tr}$ and then we create an ensemble of $M$ initial conditions $\{\bar{x}^{(k)}\}$, chosen each $T_{\text{gap}}$ time units of the simulations. There is considerable flexibility in choice of $T_{tr}$ and $T_{\text{gap}}$. $T_{tr}$ depends on the chosen initial condition and should be much larger than the time it takes for the orbit to be come in close proximity of the attractor, while $T_{\text{gap}}$ should be much longer than the inverse of the first Lyapunov exponent of the system to guarantee - at all practical levels - that the initial conditions are virtually independent. 
Our first estimate of the response function $\Gamma^+_{i,k}$ is obtained by averaging over such an ensemble of $M$ initial conditions the quantities $\delta x_{i}^{(k)}(t)$ :
\begin{equation}
    \Gamma^+_{i,k}(t)=\frac{1}{\epsilon M}\sum_{k=1}^{M}\delta x_{i}^{(k)}(t),
   \label{l8}
\end{equation}
where $\delta x_{i}^{(k)}(t)$ is the difference between the value of the variable $x_{i}$ at time $t$ in the perturbed and unperturbed run, both having the same initial condition $\bar{x}^{(k)}$. We then repeat the same procedure by switching the sign of the forcing: $\epsilon\rightarrow-\epsilon$, obtaining $\Gamma^-_{i,k}(t)$. Our estimate of the response function is then given by
\begin{equation}
    \Gamma_{i,k}(t)=\frac{\Gamma^+_{i,k}(t)+\Gamma^-_{i,k}(t)}{2},
   \label{l8b}
\end{equation}
The last step allows to remove the second order correction to the linear response and considerably increases the precision of the estimate \cite{Gritsun2017}. The linearity of the responses has been tested to hold very well up to $\epsilon=1$, see Appendix \ref{linear}. We choose $\epsilon=1$ for our numerical studies in order to have a good  signal-to-noise ratio while being within the regime of linear response. We remark that the procedure above ensures that the response function is causal.

\begin{figure}
\centering
\begin{subfigure}
  \centering
  \includegraphics[width=0.8\linewidth]{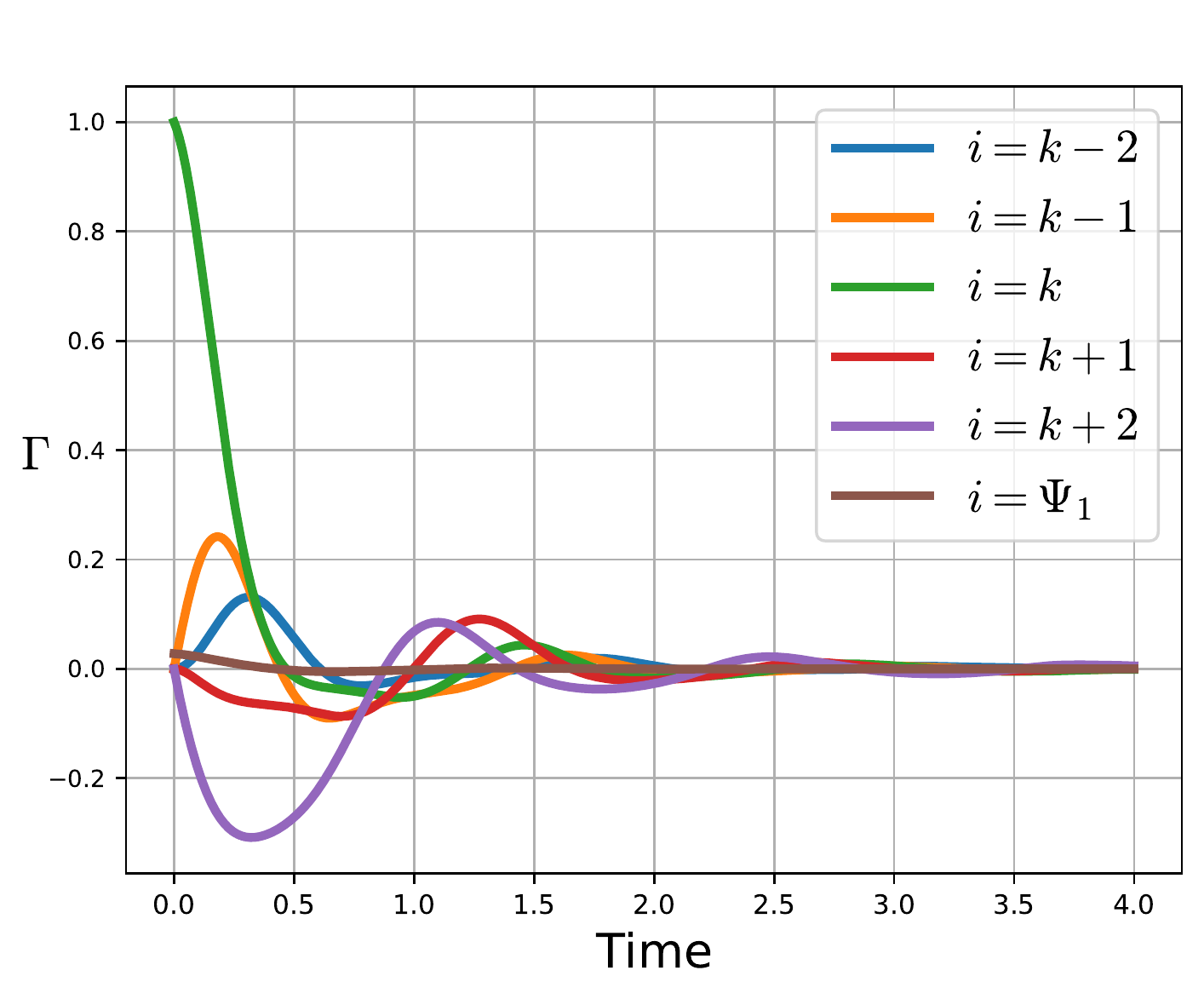}
\end{subfigure}%
\caption{Plots of the response functions $\Gamma_{i,k}$ for $i=\{k-2,k-1,k,k+1,k+2\}$ and $\epsilon=1$. We also portray $\Gamma_{\Psi_1,k}$, the response function for the mean momentum $\Psi_1$ (Eq. \ref{n200}). see also Fig. \ref{testlinearity}.}
\label{fig:gamma_functions}
\end{figure}

The response functions $\Gamma_{i,k}$ for $i=\{k-2,k-1,k,k+1,k+2\}$ obtained for $M=2\cdot 10^{6}$ and $\epsilon=1$ are shown in Fig. \ref{fig:gamma_functions}. 
We focus on the dynamical variables nearby the perturbation site because the intensity of the response decreases dramatically as we move further away, as discussed below. 

 \subsubsection{Hierarchy in predictive power} \label{detail}
 \subsubsection*{Propagation of the perturbation signal}

\begin{figure}
\centering
\begin{subfigure}
  \centering
  \includegraphics[width=0.8\linewidth]{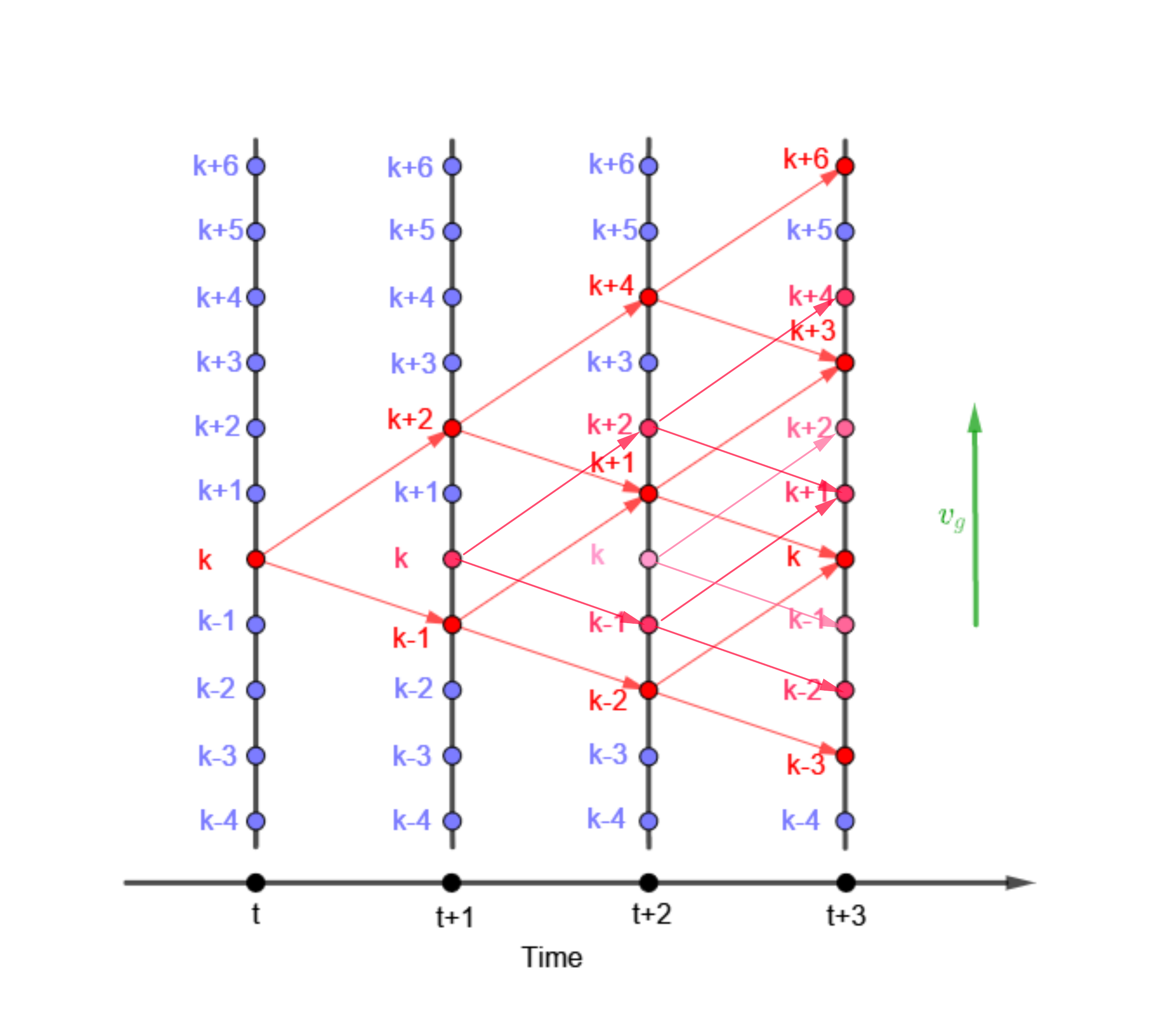}
\end{subfigure}%
\caption{Propagation of the perturbation for small times in the L96 model perturbed locally in $x_{k}$ starting from a time $t$. The vertical lines are the dynamical variable $x_{k}$ taken at different time instants, while the horizontal line below is the time axis. We have discretized the time in unit time steps for clarity purposes. At a given time, we have coloured the dynamical variables which feel directly the perturbation from the perturbed variables of the instant before with red. The colour loses its intensity as time goes by. We can notice that there are more red-coloured variables above the site $x_{k}$ than below. This is consistent with the fact that the information propagates from the dynamical variables $x_{k+j}$ with $j<0$ to the dynamical variables $x_{k+j}$ with $j>0$, with a velocity given by the group velocity $v_{g}$, shown with a green arrow in the figure.}
\label{fig:vg}
\end{figure}

We now look at the 
leading order term of short-time behaviour of $\Gamma_{i,k}$ for positive times ($\Gamma_{i,k}$ vanishes for negative times):
 \begin{equation}
     \Gamma_{i,k}(t)\underset{t\rightarrow 0^{+}}{\approx} \left\{
     \begin{aligned}
      \,&1+\mathcal{O}(t), \qquad\qquad\qquad i=k \\
      \,&\langle C^{(1)}_{i}\rangle_0 +\mathcal{O}(t^2).\,\,\,\quad i\in \{k-1,k+2\}\\
      \,& \langle C^{(q)}_{i}\rangle_0 t^{q}+\mathcal{O}(t^{q+1}),\quad i\in\{k-q,k+2q-3,k+2q\},\, q\ge2,
     \end{aligned}
     \right.
     \label{n3}
 \end{equation}
 where the coefficient $\langle C_{i}^{(q)}\rangle_0$ is the expectation value of the function $C_{i}^{(q)}$ taken over the stationary measure $\rho_0$, see Appendix \ref{asymp}. 
 In particular, we notice that at $t=0$ the response function $\Gamma_{i,k}$ is equal to $1$ if $i=k$ and it is vanishing for $i\neq k$. This is intuitive: at $t=0$ the perturbation is felt in all its intensity just at the grid point that has been directly perturbed. As $t$ increases, the perturbation propagates also to the other grid points $x_{i}$, with a time scale determined by the leading order term given in  Eq. \ref{n3}.  Note that the perturbation propagates more efficiently towards the right ($i>k$) than towards left ($i<k$), since for each dynamical variable $x_{k-q}$ at the left of $x_{k}$ with leading term $t^{q}$ there are two dynamical variables $x_{k+2q-3}$ and $x_{k+2q}$ (for $q>1$) at its right with the same leading term. We can have a clearer intuition of the actual propagation of the signal in space and in time by looking at the cartoon in Fig. \ref{fig:vg}. It is remarkable that the site $k+3$ or $k+5$ are affected by the forcing later than the sites $k+4$ or $k+6$, respectively. This further indicates that the advection taking place in the L96 system is a non-standard one. 

The high-frequency asymptotic behaviour of the susceptibilities corresponding to the response functions given in Eq. \ref{n3} can be derived using Eq. \ref{e100}. 

\subsubsection*{Hierarchy of the dynamical variables}

\begin{figure}
\centering
\begin{subfigure}
  \centering
  \includegraphics[width=0.8\linewidth]{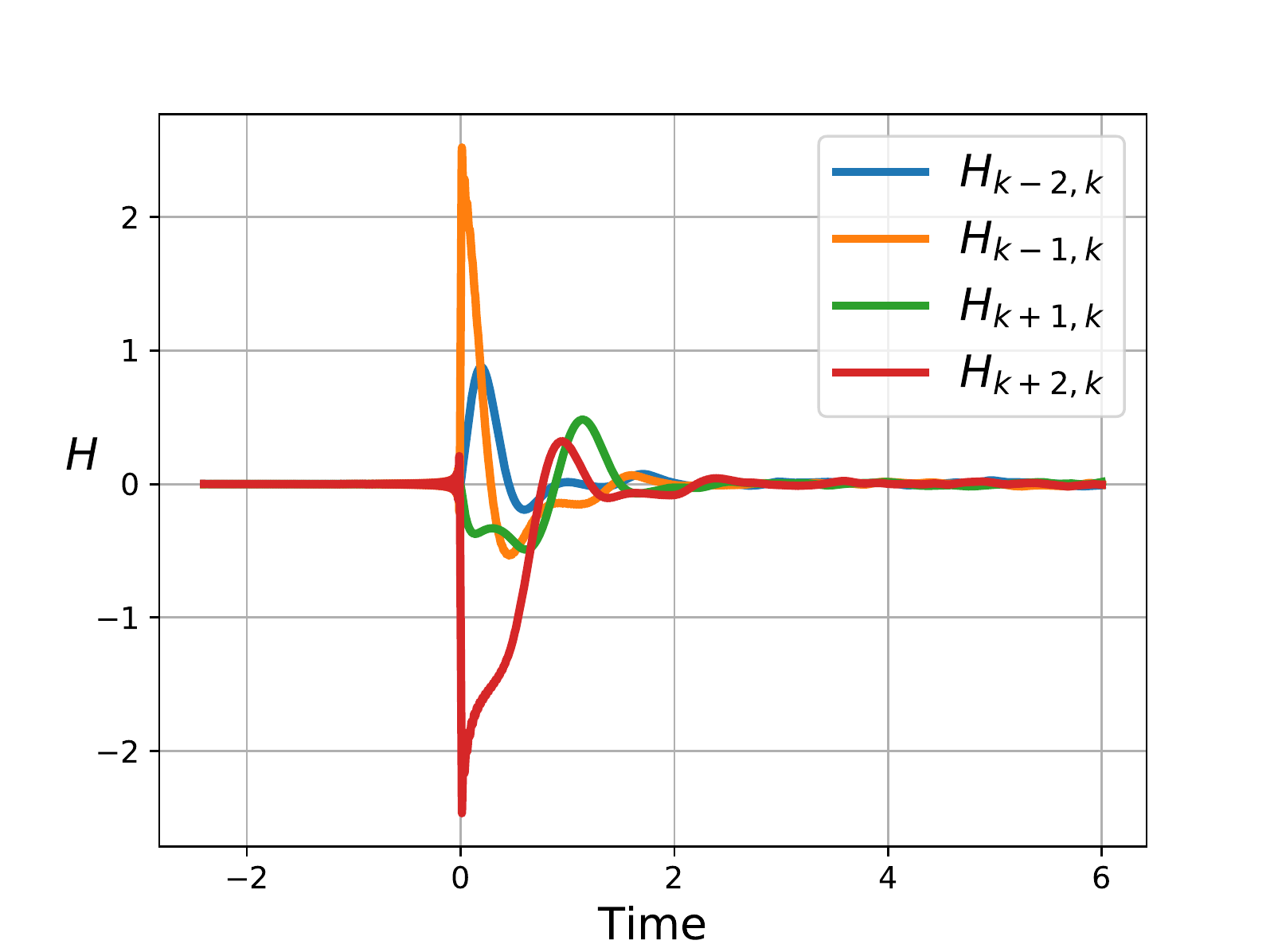}
\end{subfigure}%
\caption{Surrogate response functions $H_{i,k}$ for $i\in\{k-2,k-1,k,k+1,k+2\}$.}
\label{fig:h18}
\end{figure}
\begin{figure}
\centering
\begin{subfigure}
  \centering
a)  \includegraphics[width=0.45\linewidth]{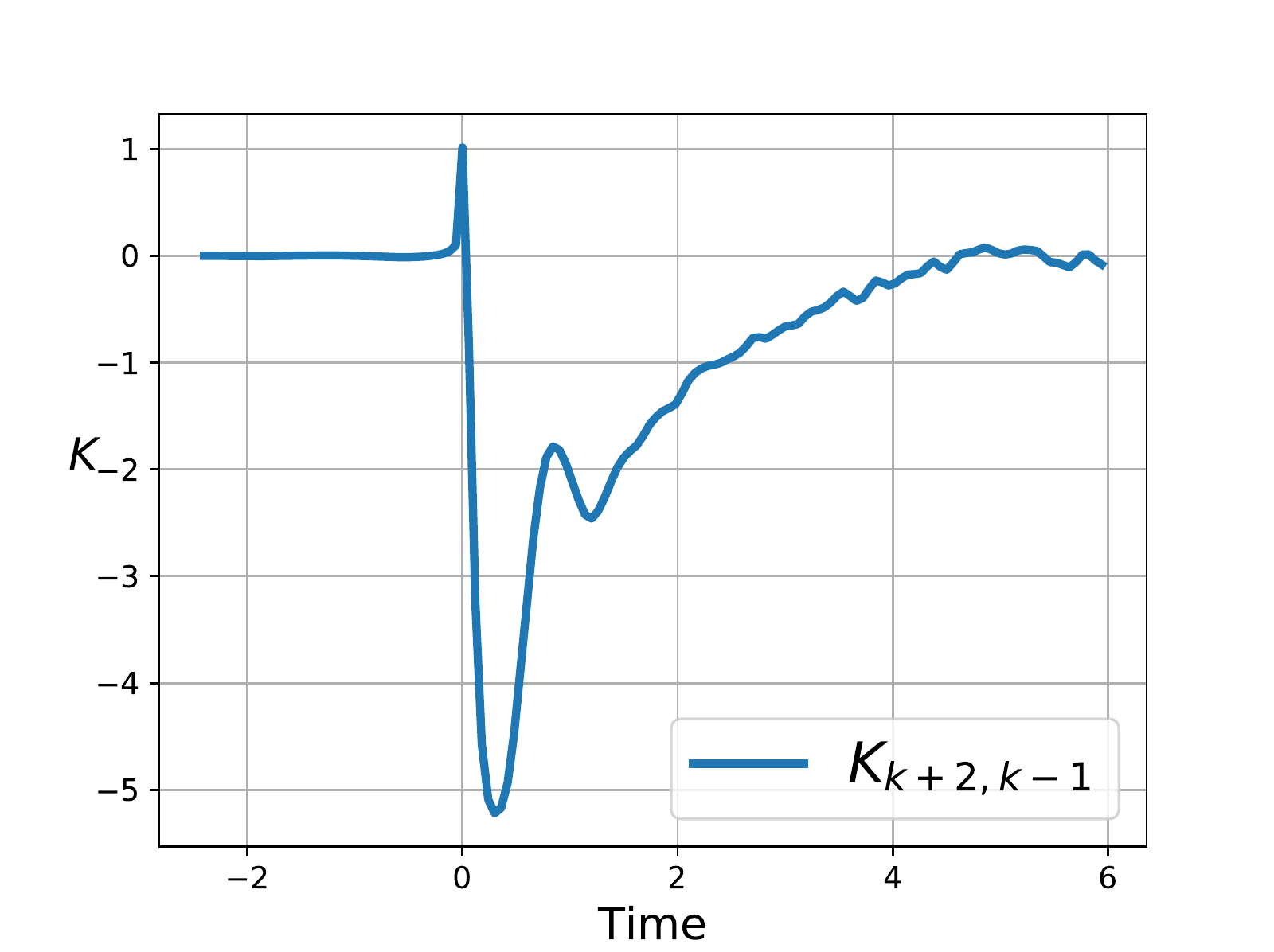}
\end{subfigure}%
\begin{subfigure}
  \centering
b)  \includegraphics[width=0.45\linewidth]{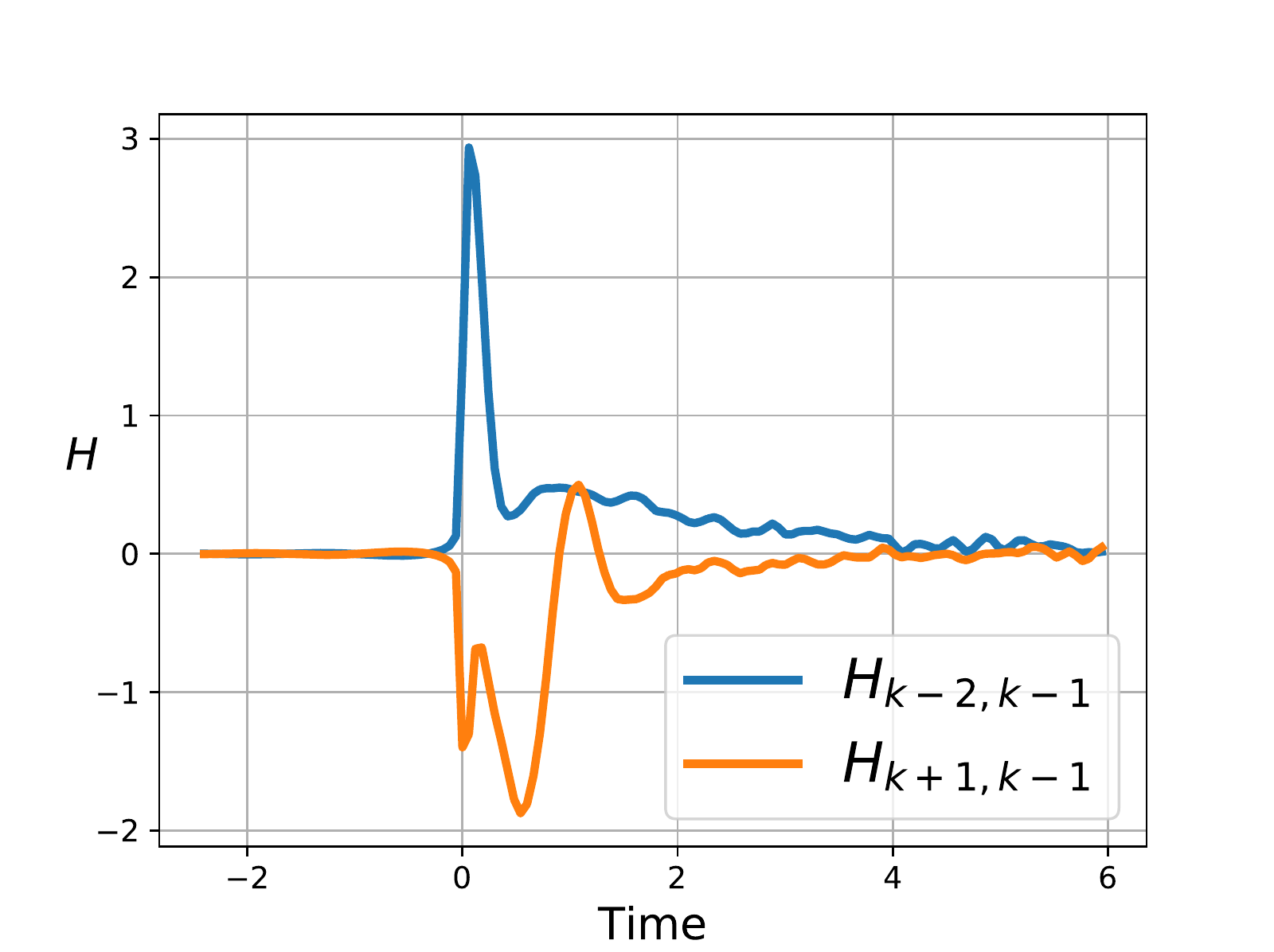}
\end{subfigure}
\caption{ (a) Non-singular part of the surrogate response function $K_{k+2,k-1}$. (b) Surrogate response functions $H_{k-2,k-1}$ and $H_{k+1,k-1}$.}
\label{fig:H_17}
\end{figure}
\begin{figure}
\centering
\begin{subfigure}
  \centering
a)  \includegraphics[width=0.45\linewidth]{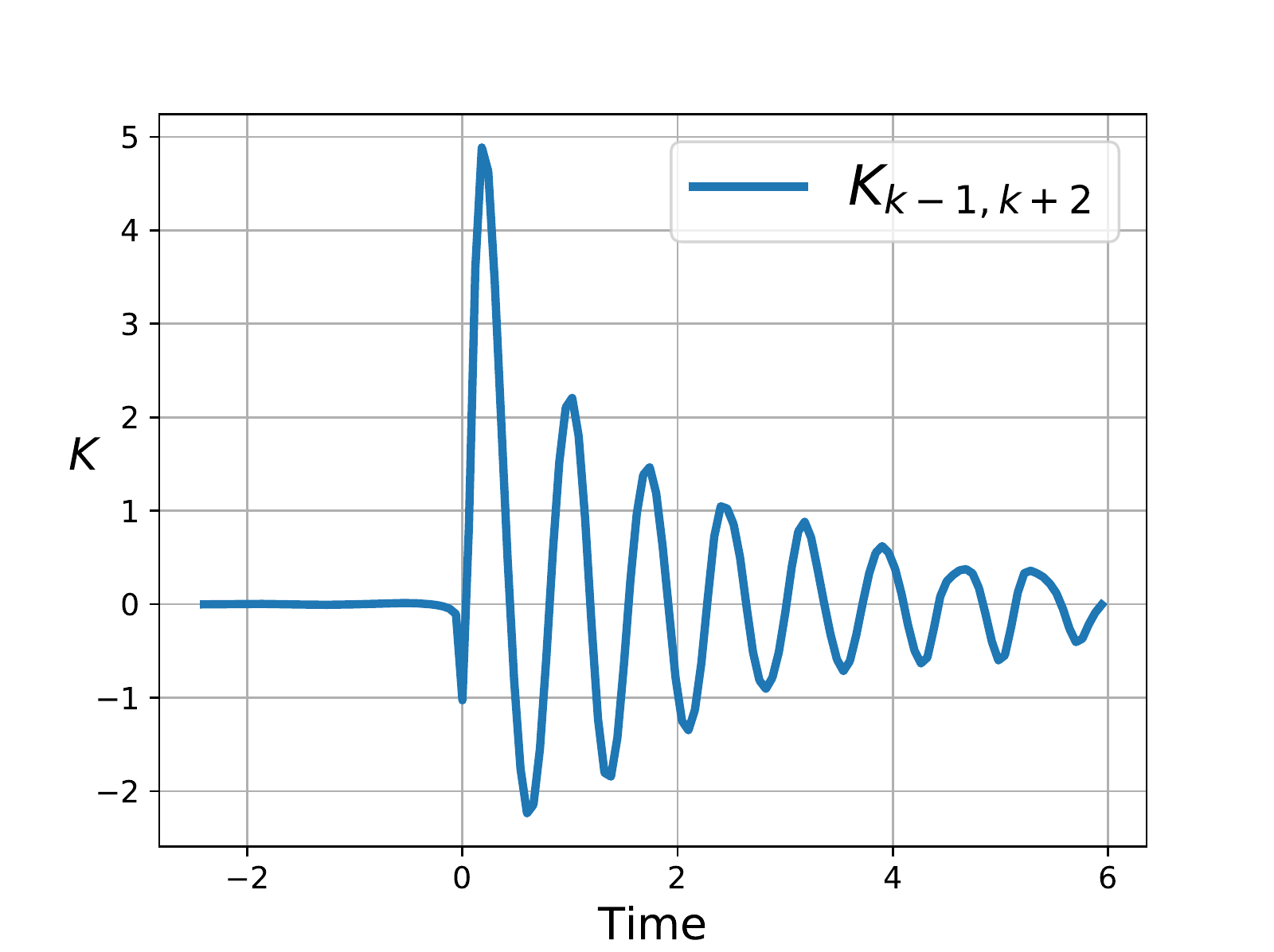}
\end{subfigure}%
\begin{subfigure}
  \centering
b)  \includegraphics[width=0.45\linewidth]{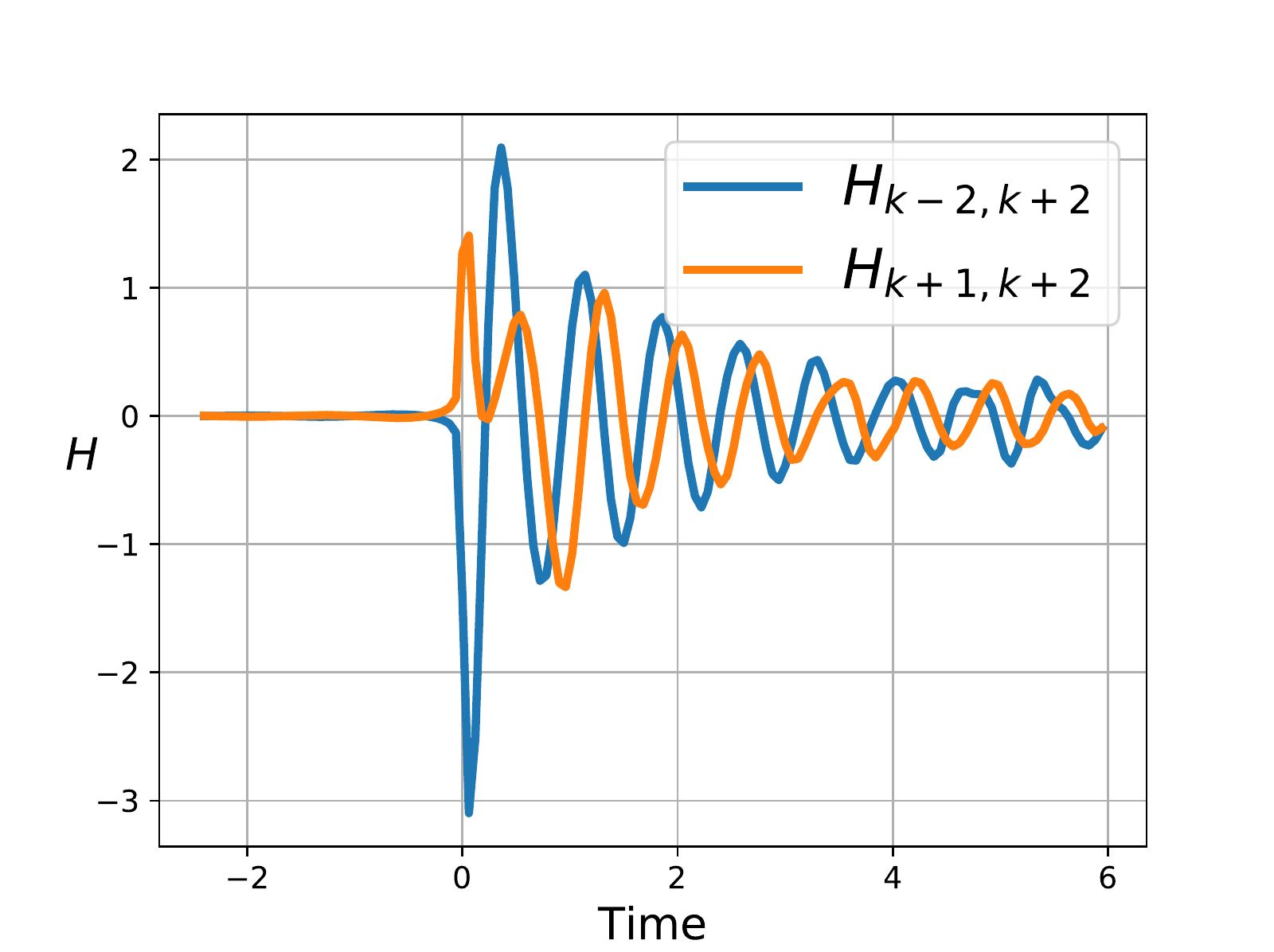}
\end{subfigure}
\caption{ (a) Non-singular part of the surrogate response function $K_{k-1,k+2}$. (b) Surrogate response functions $H_{k-2,k+2}$ and $H_{k+1,k+2}$.}
\label{fig:H_20}
\end{figure}
\begin{figure}
\centering
\begin{subfigure}
  \centering
 a) \includegraphics[width=0.45\linewidth]{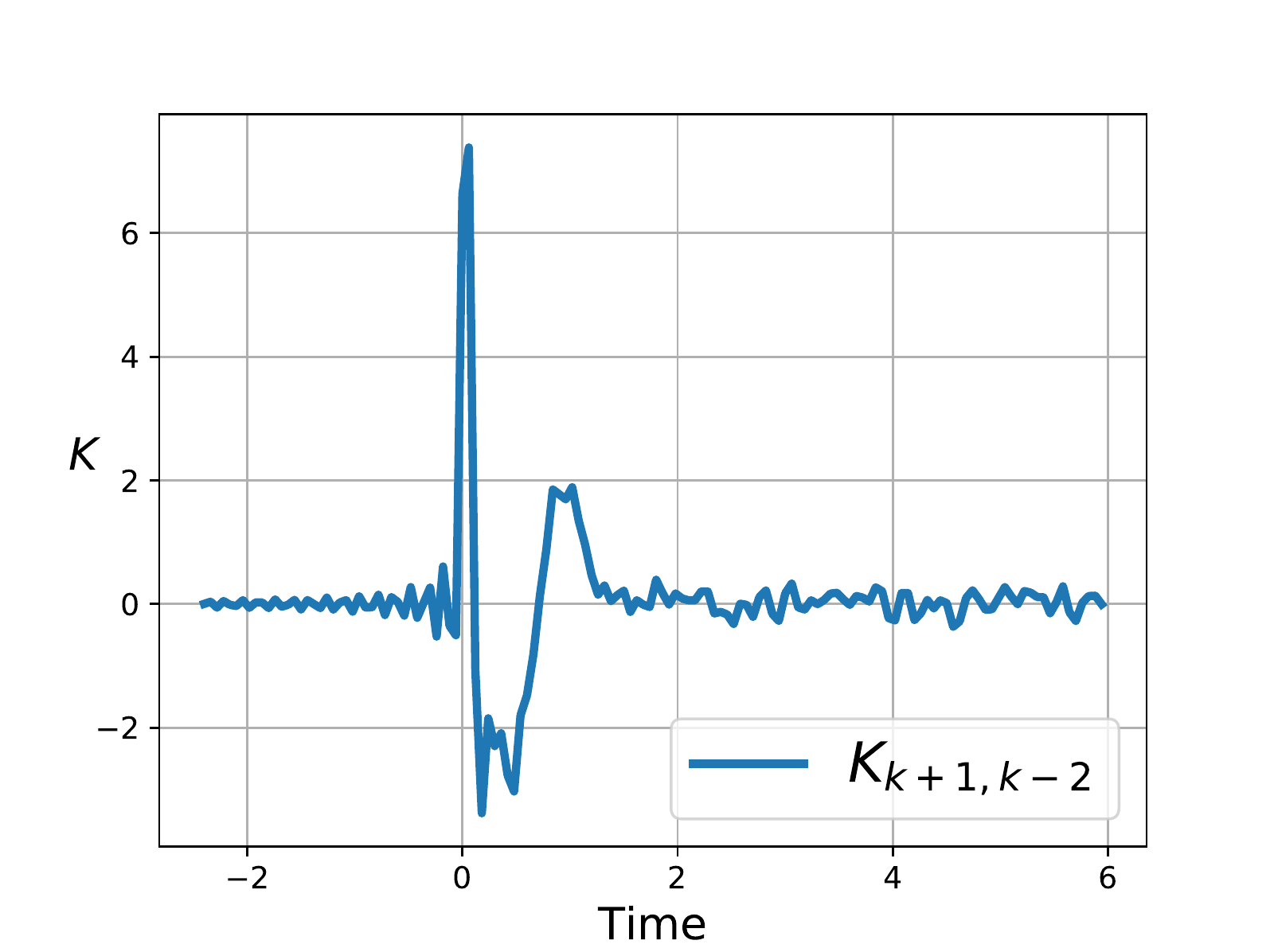}
\end{subfigure}%
\begin{subfigure}
  \centering
  b)\includegraphics[width=0.45\linewidth]{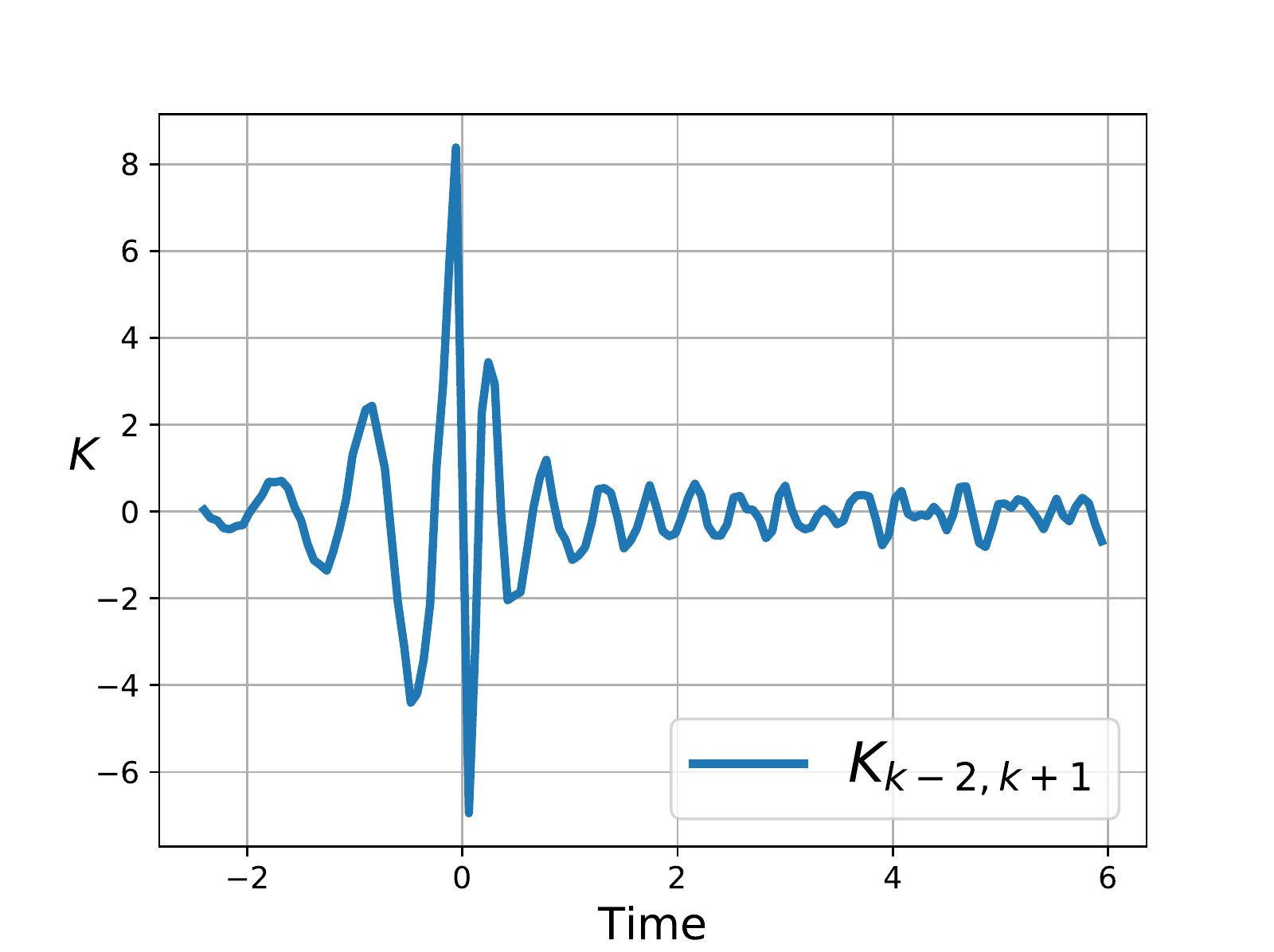}
\end{subfigure}
\caption{ (a) Non-singular part of the surrogate response function $K_{k+1,k-2}$. (b) Non-singular part of the surrogate response function $K_{k-2,k+1}$.}
\label{fig:H_1619}
\end{figure}

We want now to build 
a hierarchy of the dynamical variables  in terms of their predictive power of the other dynamical variables. In particular, this hierarchy is closely related to how the perturbation signal propagates in the system: the sooner a dynamical variables feels the perturbation and the more predictive it will be. We construct the surrogate response function by using Eq. \ref {s3} and by then applying the cutoff introduced in Eq. \ref{0f}. We define as $H_{i,j}$ the surrogate response function that allows one to reconstruct the variable $i$ using the variable $j$ as surrogate forcing. We indicate with $K_{i,j}$ ($S_{i,j}$) its non-singular (singular) component, and with $H^c_{i,j}$ its causal component,   

The variable with the highest predictive power is obviously $x_{k}$, since it responds immediately to the perturbation applied at the site $k$. Indeed, the non-causal component of the surrogate response functions shown in Fig. \ref{fig:h18} is entirely negligible. The second most predictive variables are $x_{k-1}$ and $x_{k+2}$. A few surrogate response functions employing them as predictors are shown in Fig. \ref{fig:H_17} and Fig. \ref{fig:H_20}. We can observe that these surrogate response functions show a rather small non-causal component. Notice that just the non-singular components Eq. \ref{w1} are shown in the figures, for clarity purposes. The third most predictive variables between the ones we have considered are $x_{k-2}$ and $x_{k+1}$. In Fig. \ref{fig:H_1619} we show the non-singular parts of the related surrogate response functions $K_{k-2,k+1}$ and $K_{k+1,k-2}$. We can notice that their non-causal components are more relevant than the ones considered before, because the information retained by the predictors is degraded. 



 We can better quantify the importance of the non causal component of the surrogate response functions through the PI introduced in Eq. \ref{t6}. 
 The results are presented in Table \ref{table:method} for the surrogate response functions between the dynamical variables $x_{k-2}$, $x_{k-1}$, $x_{k+1}$ and $x_{k+2}$. Looking at the values provided by the PI, we observe that the weight of the non causal part is bigger when the predictors are $x_{k-2}$ and $x_{k+1}$. This is in agreement with the hierarchy described above.  
 
 
 Another aspect we wish to analyse is whether one can define a preferential direction for performing the prediction. If we take two variables $x_{i}$ and $x_{j}$, we can ask ourselves whether it is better to use $x_{i}$ to predict $x_{j}$ or the other way around. Of course, this issue makes sense in the case the two variables $x_{i}$ and $x_{j}$ have the same rank (otherwise we would just use the higher ranked variable as a predictor). This is the case of $x_{k-1}$ and $x_{k+2}$ and of $x_{k-2}$ and $x_{k+1}$. Making use of the table \ref{table:method}, we see that in both cases the variable with lower index (more to the left) is the better predictor. This is consistent with the fact that the group velocity $v_{g}$ of the travelling waves in L96, which controls the flow of information, is positive (from left to right): the variables that are situated upstream predict better than those situated downstream. 
\begin{table}
\begin{centering}
 \begin{tabular}{c|cccc}
	\toprule
	\,&  $x_{k-2}$&  $x_{k-1}$&  $x_{k+1}$&  $x_{k+2}$\\ 
	 \midrule
	$x_{k-2}$& $\cdot$ &  $\cdot$&  0.083 & $\cdot$ \\ 
	 
	$x_{k-1}$&  0.013 & $\cdot$ &  0.019& 0.0022 \\ 
	$x_{k+1}$& 0.58 & $\cdot$ & $\cdot$ &$\cdot$  \\ 
	$x_{k+2}$& 0.0068 &  0.0033 &  0.014& $\cdot$ \\ 
	\bottomrule
\end{tabular}
\caption{PI for the surrogate response functions between the dynamical variables $x_{k-2}$, $x_{k-1}$, $x_{k+1}$ and $x_{k+2}$. In the first column there are the predictors, while in the first row there are the predictands.}
\label{table:method}
\end{centering}
\end{table} 

	 


%
%
%

\begin{figure}
\centering
\begin{subfigure}
  \centering
  \includegraphics[width=0.8\linewidth]{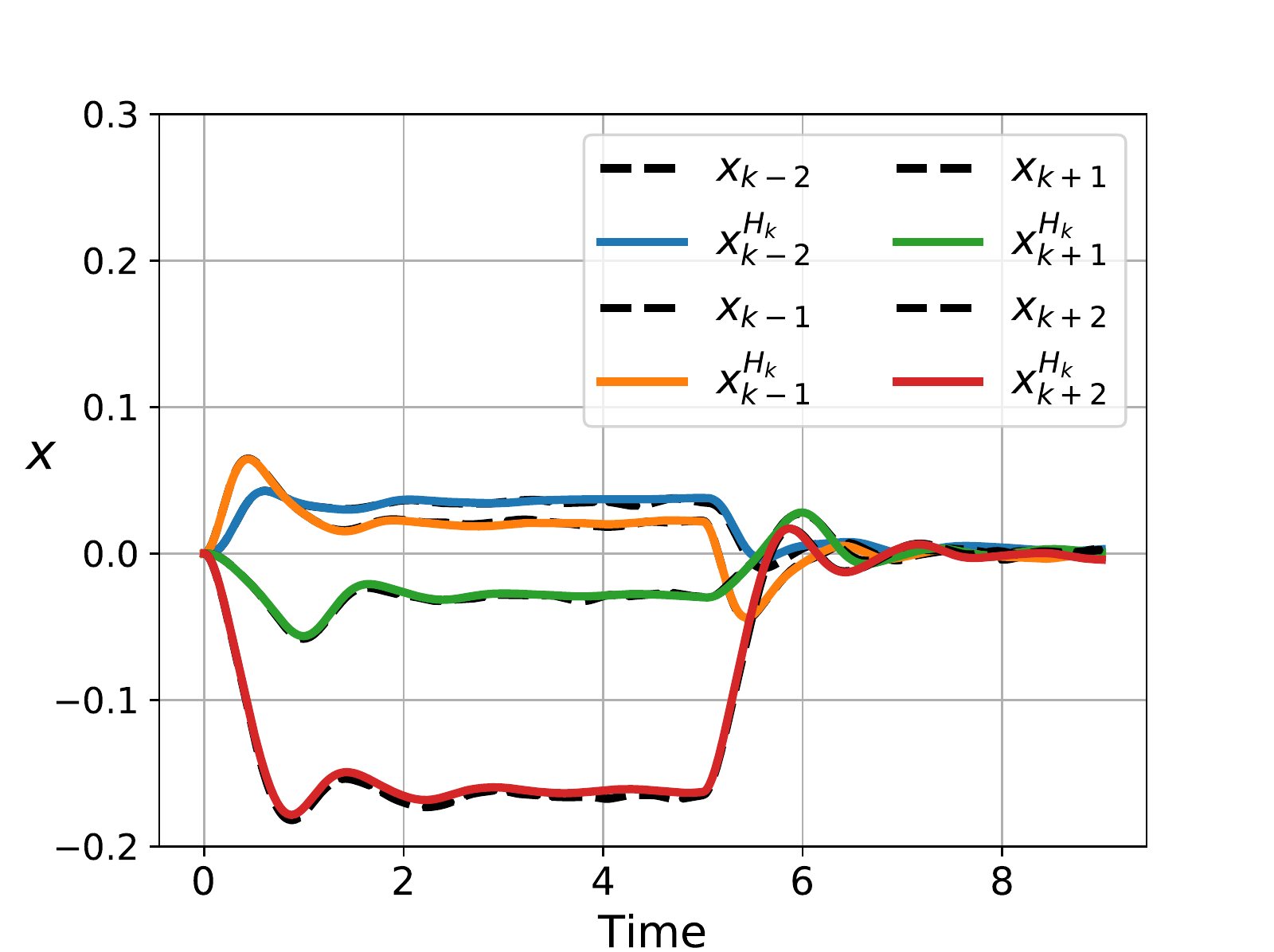}
\end{subfigure}%
\caption{Comparison between the response of the perturbed L96 system to perturbation with spatial pattern Eq. \ref{l7} and time pattern Eq. \ref{n36} with $\tau=5$ and the predictions made using the surrogate response functions $H_{i,k}$ for $i\in\{k-2,k-1,k+1,k+2\}$.}
\label{fig:pred_18}
\end{figure}

Now we test the actual predictive ability of the surrogate response function computed above. We perturb the L96 system with the vector field having 
spatial pattern $G(x)=\epsilon\delta_{i,k}$ as above and having the following time pattern:
\begin{equation}
e(t)=\Theta(t)-\Theta(t-\tau),
\label{n36}
\end{equation}
with $\tau=5$. This time pattern seems relevant because corresponds to the  act of switching abruptly on and off the perturbation, and keeping it active for a time scale that is longer than the inverse of the first Lyapunov exponent of the system (about 0.6 time units). Therefore, it allows to appreciate both transient and longer term features of the response of the system.  We then test the skill of the  variable $x_{k}$ in predicting $x_{j}$, with $j\in\{k-2,k-1,k+1,k+2\}$:
\begin{equation}
    \delta \langle x_{i} \rangle (t) =\int_{-\infty}^{\infty}d\tau\, H^{c}_{ij}(t-\tau) \delta\langle x_{j} \rangle (\tau),
    \label{n40}
\end{equation}
where we use only the causal component of the surrogate response function. 
The predictions are shown in Fig. \ref{fig:pred_18}, where we can notice that the agreement between the actual response and the prediction made through the surrogate response functions where $x_{k}$ is the predictor is very good in all cases.

\begin{figure}
\centering
\begin{subfigure}
  \centering
a)  \includegraphics[width=0.45\linewidth]{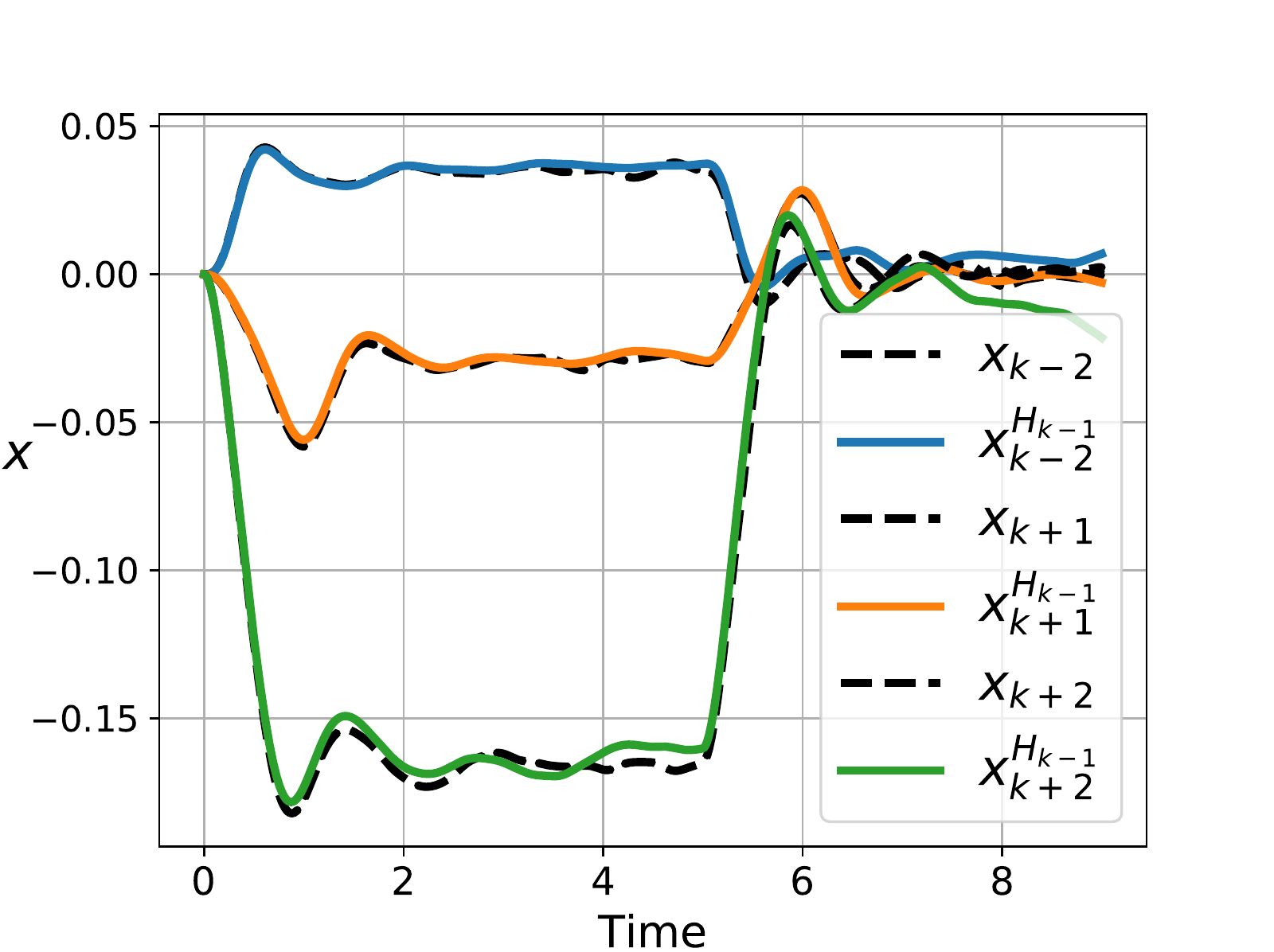}
\end{subfigure}
\begin{subfigure}
  \centering
b) \includegraphics[width=0.45\linewidth]{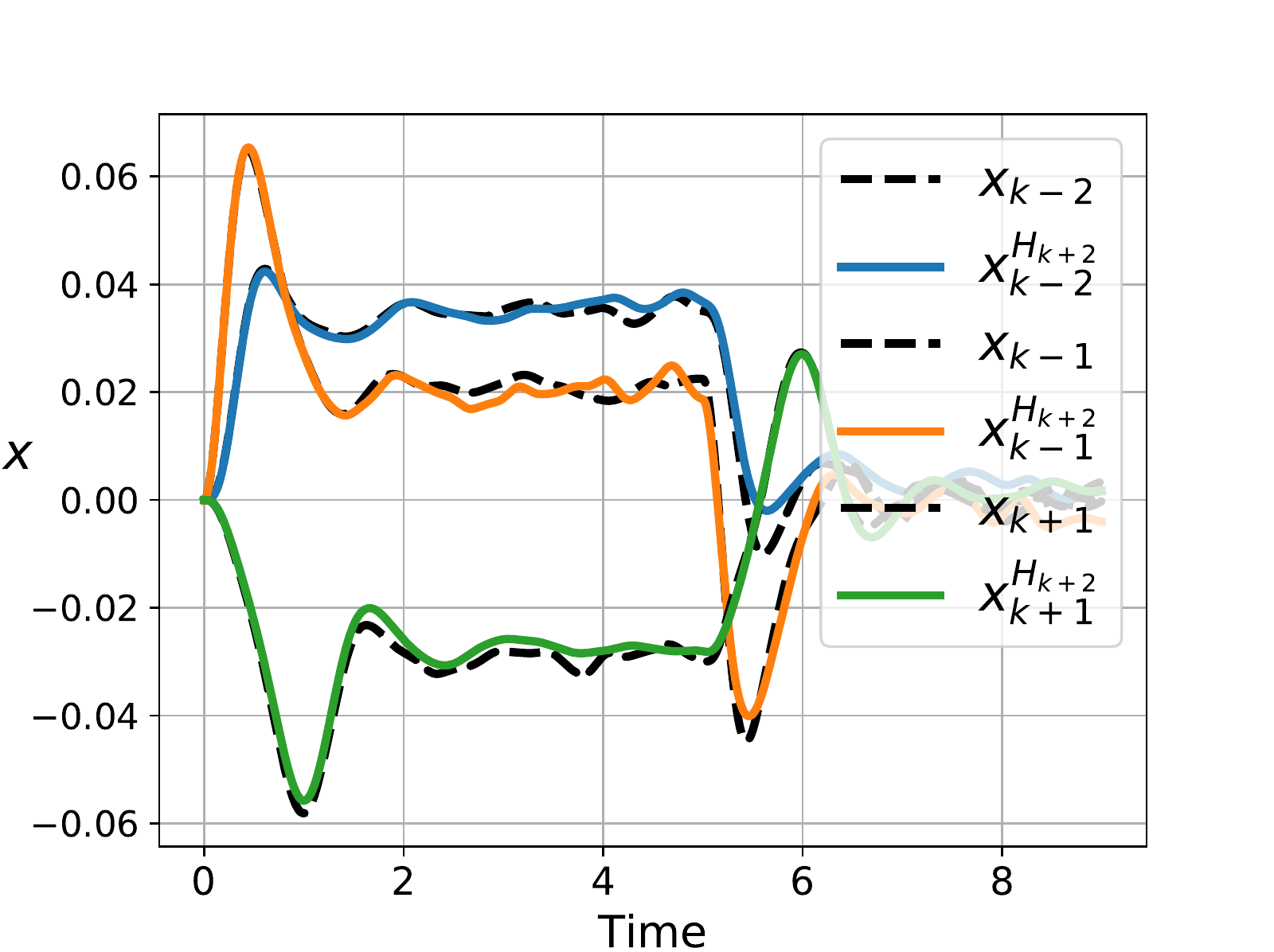}
\end{subfigure}
\caption{Same as Fig. \ref{fig:pred_18}, but looking at predictions performed using $x_{k-1}$ in (a) and with  $x_{k+2}$ in (b).}
\label{fig:pred_1720}
\end{figure}

We now consider the second most predictive dynamical variables $x_{k-1}$ and $x_{k+2}$. The predictions are shown in Fig. \ref{fig:pred_1720}. We can notice that these predictors work quite well: despite not being directly perturbed by the forcing, they retain a lot of information. We also remark that some discrepancy between prediction and the actual response emerges in terms of mismatch of the oscillations taking place during the plateau of the forcing. 

\begin{figure}
\centering
\begin{subfigure}
  \centering
a)  \includegraphics[width=0.45\linewidth]{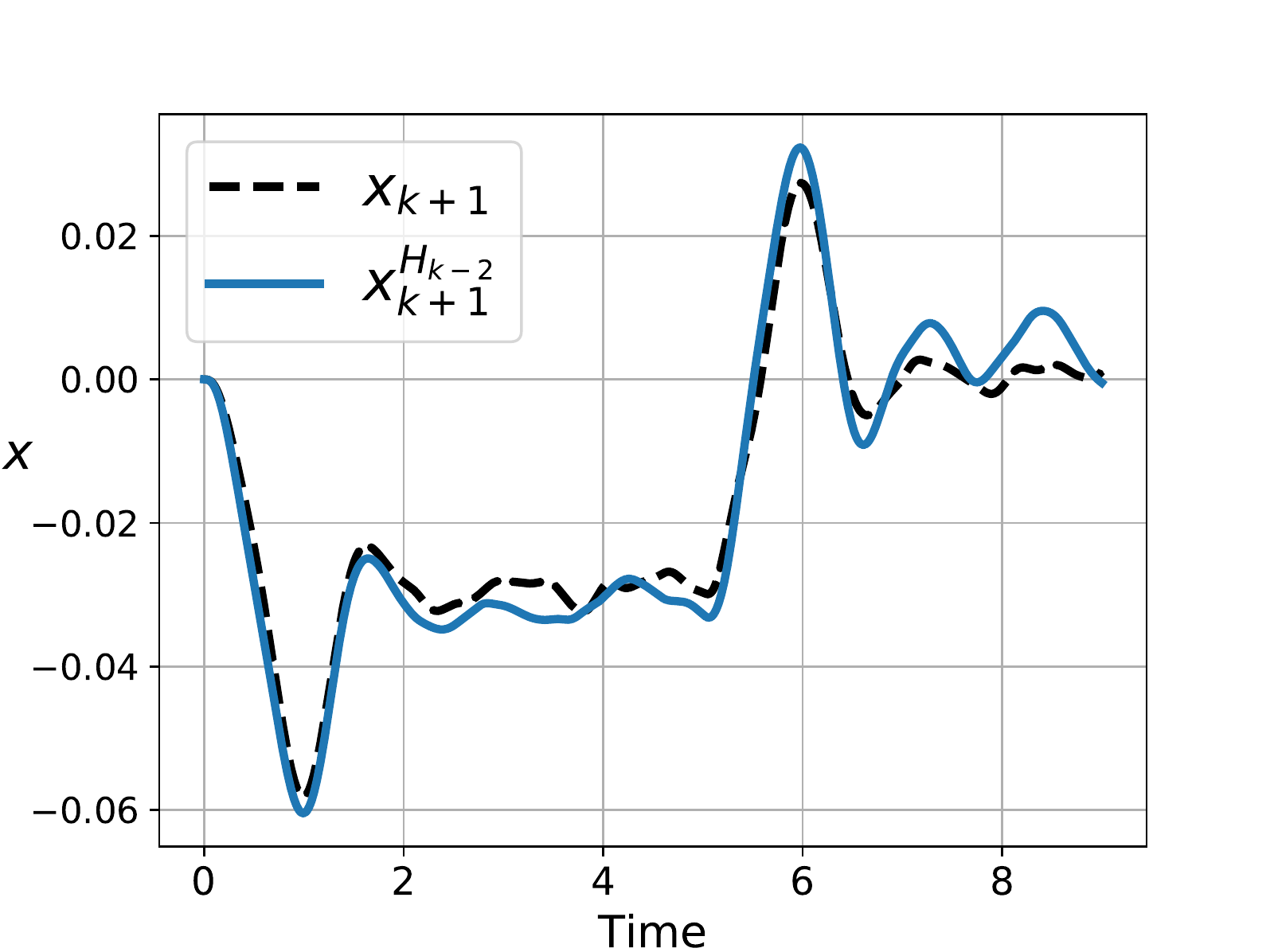}
\end{subfigure}%
\begin{subfigure}
  \centering
b)  \includegraphics[width=0.45\linewidth]{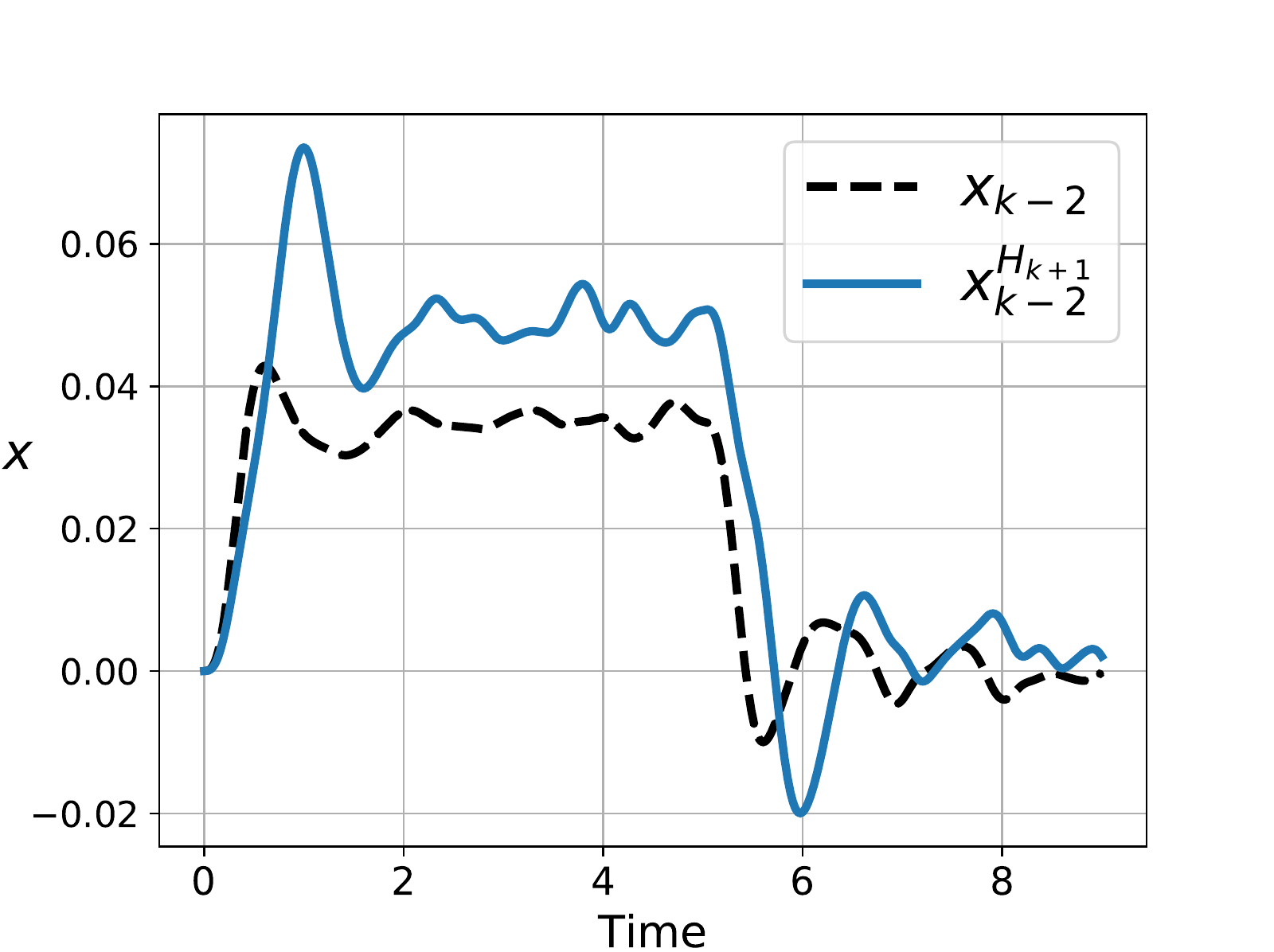}
\end{subfigure}
\caption{Same as Fig. \ref{fig:pred_18}, but looking at predictions performed using $x_{k-2}$ in (a) and with  $x_{k+1}$ in (b).}
\label{fig:pred_1619}
\end{figure}

Lastly, we take into account the predictions made using the variables $x_{k-2}$ and $x_{k+1}$. As we can see in Fig. \ref{fig:pred_1619}, the predictions are clearly less successful than in previous cases, even though they show a qualitative agreement with the actual responses. This is explained by the greater relevance of the non-causal components of the related surrogate response functions $H_{k-2,k+1}$ and $H_{k+1,k-2}$, as it can be seen in Fig. \ref{fig:H_1619}; see also  Tab. \ref{table:method}. Remarkably, be comparing the two panels of Fig. \ref{fig:pred_1619}, we can clearly see the asymmetry between the mutual predictive power of $x_{k-2}$ (better) and $x_{k+1}$ (worse)  discussed above. 



\subsection{Making predictions with more observables}\label{more}
We have shown above that some local variables cannot well predict other local variables, as a result of how the perturbation signal propagates across the system. 
Following the discussion in Sect. \ref{srt}, we test whether this can be improved by applying two independent forcings and, correspondingly, using two predictors. The idea is that by adding a forcing and a predictor we are able to learn more about the properties of the system and of its response. We then add, on top of the perturbation described by Eq. \ref{l7} with time pattern given by Eq. \ref{n36},
a second forcing that impacts the viscosity of the system acting at the $k^{th}$ grid point : 
\begin{equation}
    G_{i}^{(2)}(x)= -x_{i}\,\delta_{ik}\epsilon_{2}.
    \label{n206}
\end{equation}
The forcing is applied using the temporal pattern $$e^{(2)}(t)=\Theta(t)-\Theta(t-\tau_{2}),$$ where $\tau_{2}=3$. 

\begin{figure}
\centering
\begin{subfigure}
  \centering
  \includegraphics[width=0.8\linewidth]{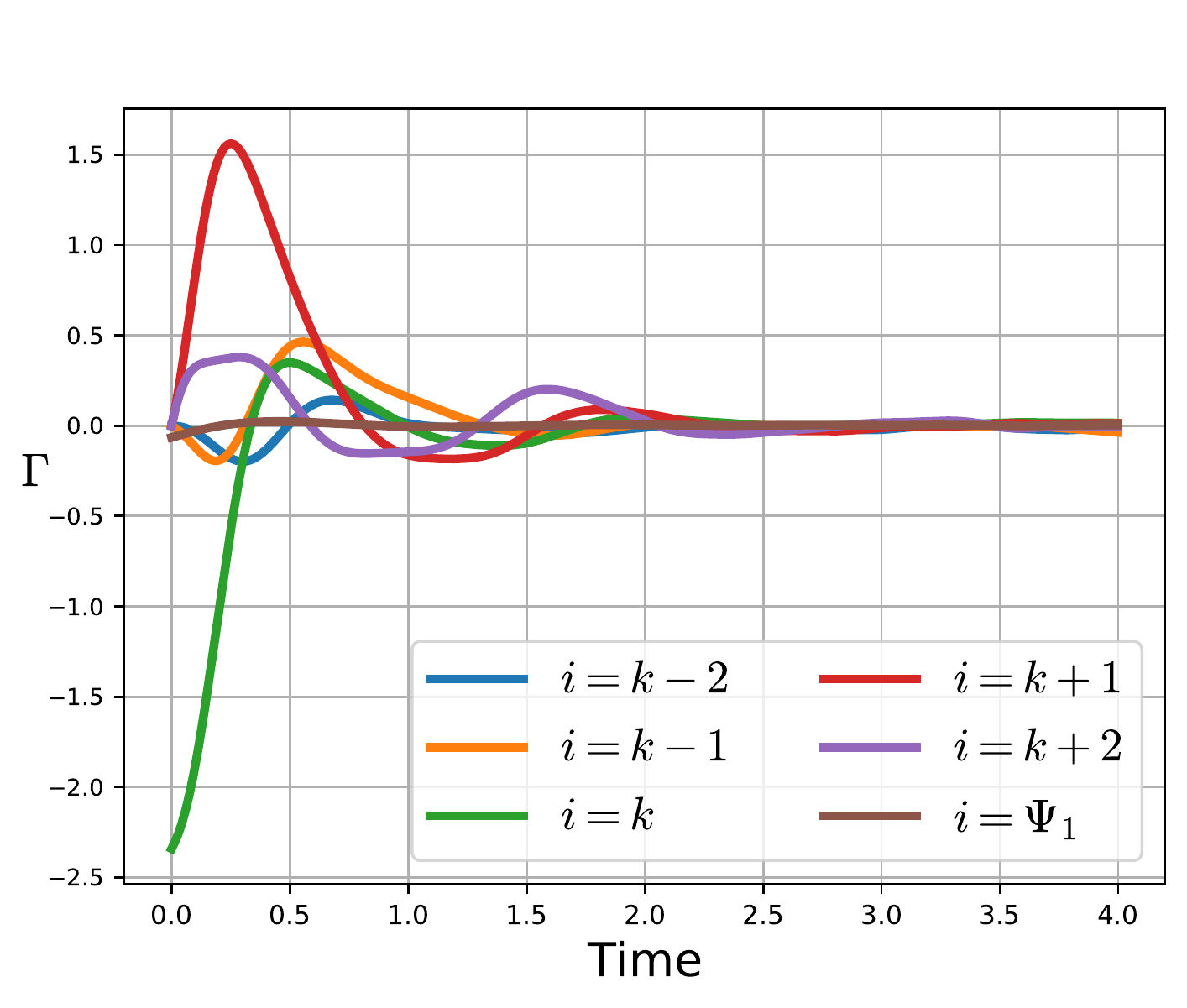}
\end{subfigure}%
\caption{Plots of the response functions $\Gamma^{(2)}_{i,k}$ for $i=\{k-2,k-1,k,k+1,k+2\}$ and $\epsilon=0.1$. We also portray $\Gamma^{(2)}_{\Psi_1,k}$, the response function for the mean momentum $\Psi_1$ (Eq. \ref{n200}). See also Fig. \ref{testlinearity2}.}
\label{fig:gamma_functions2}
\end{figure}

The corresponding response functions $\Gamma^{(2)}_{i,k}$ for $i=\{k-2,k-1,k,k+1,k+2\}$ obtained for $M=2\cdot 10^{6}$ and $\epsilon=0.1$ are shown in Fig. \ref{fig:gamma_functions2}. 
We have tested the linearity of the response of the system for values of $\epsilon_2$ ranging from 0.01 up to 2 and found that nonlinear corrections are negligible for $\epsilon_2\leq0.25$; see Appendix \ref{linear}. Note that here we need to consider smaller values of $\epsilon_2$ compared to the case of the forcing  $G^{(1)}$ because of the presence of the factor $x_i$ ($|x_i|$ is typically larger than one in the unperturbed runs).  We then perform the data analysis for the case of the combined forcings $G^{(1)}$ and $G^{(2)}$ using $\epsilon_{2}=0.1$. 

\begin{figure}
\centering
  \includegraphics[width=.8\linewidth]{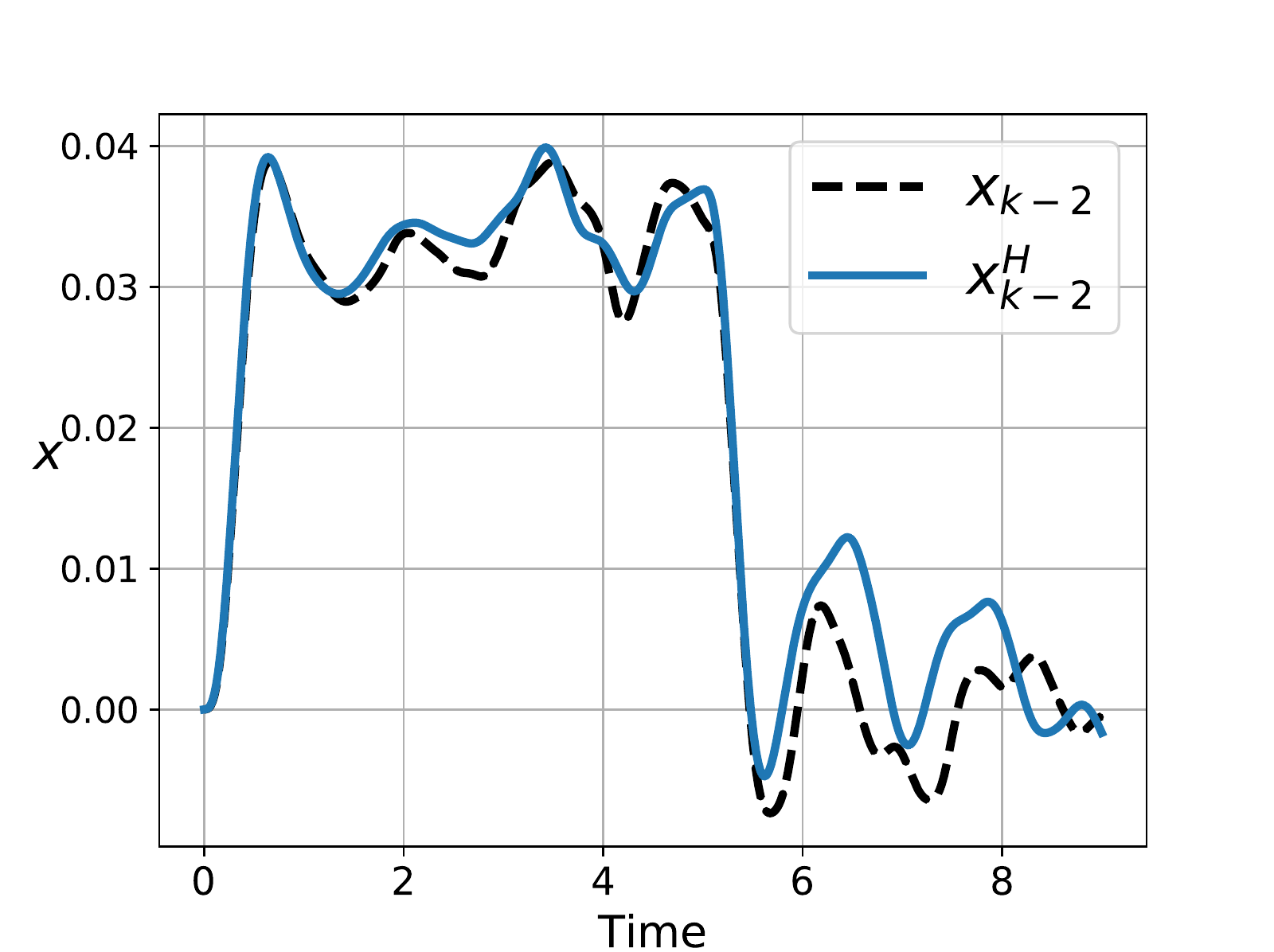}
\caption{Response of $x_{k-2}$ to two acting forcings (dashed line) and prediction obtained using $x_{k+1}$ and $\Psi_1$ as predictors.}
\label{fig:conv_impr}
\end{figure}

As shown in Eq. \ref{s19} for $N=2$, it is then necessary to choose a second observable as predictor. We choose the following global observable:
\begin{equation}
    \Psi_{1}(t)\equiv\frac{1}{N}\sum_{i=1}^{N}x_{i}.
    \label{n200}
\end{equation}
Note that $\Psi_1$ is usually interpreted as the mean momentum of the system \cite{luc11}. The corresponding response function  $\Gamma_{\Psi_1,k}(t)$ ($\Gamma^{(2)}_{\Psi_1,k}(t)$) to the forcing defined in Eq. \ref{l7} (Eq. \ref{n206}) is portrayed in Fig. \ref{fig:gamma_functions} (Fig. \ref{fig:gamma_functions2}). By symmetry, $\Gamma_{\Psi_1,k}(t)$ and $\Gamma^{(2)}_{\Psi_1,k}(t)$ do not depend on $k$. This choice is motivated by the fact that we wish to simulate the situation where a local observer, e.g. $x_{k+1}$, uses information on its own local state and on global properties of the system to predict the state of another local observer, e.g. $x_{k-2}$. We perform the prediction using the following formula: 
\begin{equation}
  \delta\langle x_{k-2}\rangle(t)=H^{c}_{k-2,\Psi_{1}}(t)\ast\delta\langle\Psi_{1}\rangle(t)+  H^{c}_{k-2,k+1}(t)\ast\delta\langle x_{k+1}\rangle(t).
\label{n203}
\end{equation}
These surrogate response functions have in general different poles than the ones previously defined in the case of just one forcing, see discussion in Appendix \ref{sing_one}. 
In Fig. \ref{fig:more16} the surrogate response function $H_{k-1,\Psi_{1}}$ and the non-singular component $K_{k-2,k+1}$ of the surrogate response function $H_{k-2,k+1}$ are shown. By comparing Figs. \ref{fig:H_1619}b and \ref{fig:more16}b, we note that the non-causal component of the response function associated to $x_{k-2}$ is greatly reduced once a second forcing and a second observable are used. Additionally, we also note that the amplitude of the surrogate response function associated to $x_{k-2}$ is greatly reduced, implying that most of the information on $x_{k-2}$ is drawn from $\Psi_1$ in the case analysed here. The improvement in our ability to predict $x_{k-2}$ is summarised in Table \ref{tab:improvement}.  By comparing the dashed lines in Fig. \ref{fig:conv_impr} and in Fig. \ref{fig:pred_1619}, one notices that adding the second forcing given in Eq. \ref{n206} has little effect on the actual response of $x_{k-2}$; indeed, the contribution to the response is smaller by a factor $\mathcal{O}(10^{-2})$ (not shown) with respect to what coming from the forcing given in Eq. \ref{l7}. But, instead, our ability to predict using surrogate response functions increases substantially when we add $\Psi_1$ as predictor, even if such observable has little information on the local properties of the system. This is due to the fact that adding a second perturbation to the system and a second observable as predictor regularises our problem. Indeed, a greater predictive skill is obtained even if we perform simulations with smaller values of $\epsilon_2$ than what reported above (not shown).


\begin{figure}
\centering
\begin{subfigure}
  \centering
a)  \includegraphics[width=0.42\linewidth]{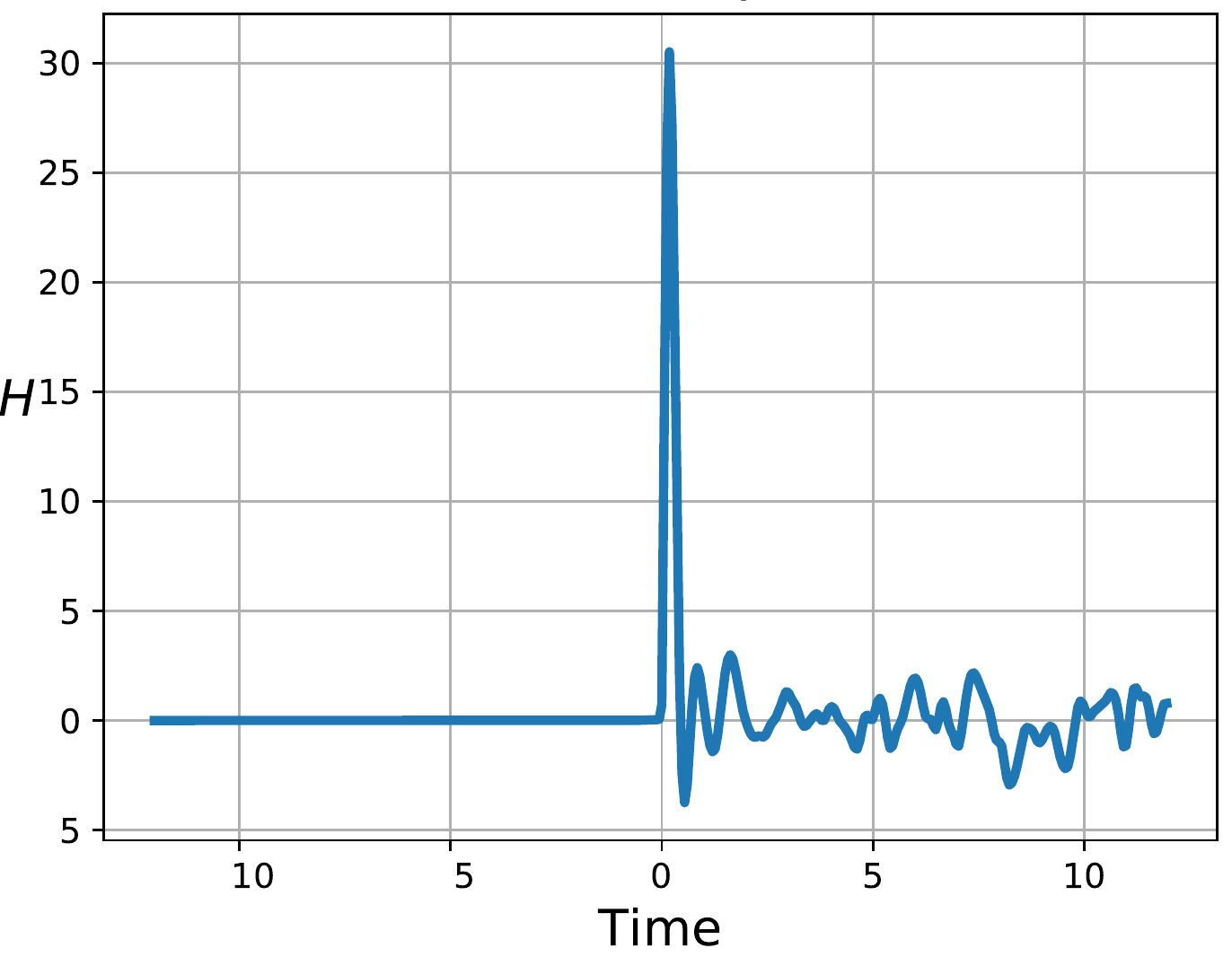}
\end{subfigure}%
\begin{subfigure}
  \centering
b)  \includegraphics[width=0.49\linewidth]{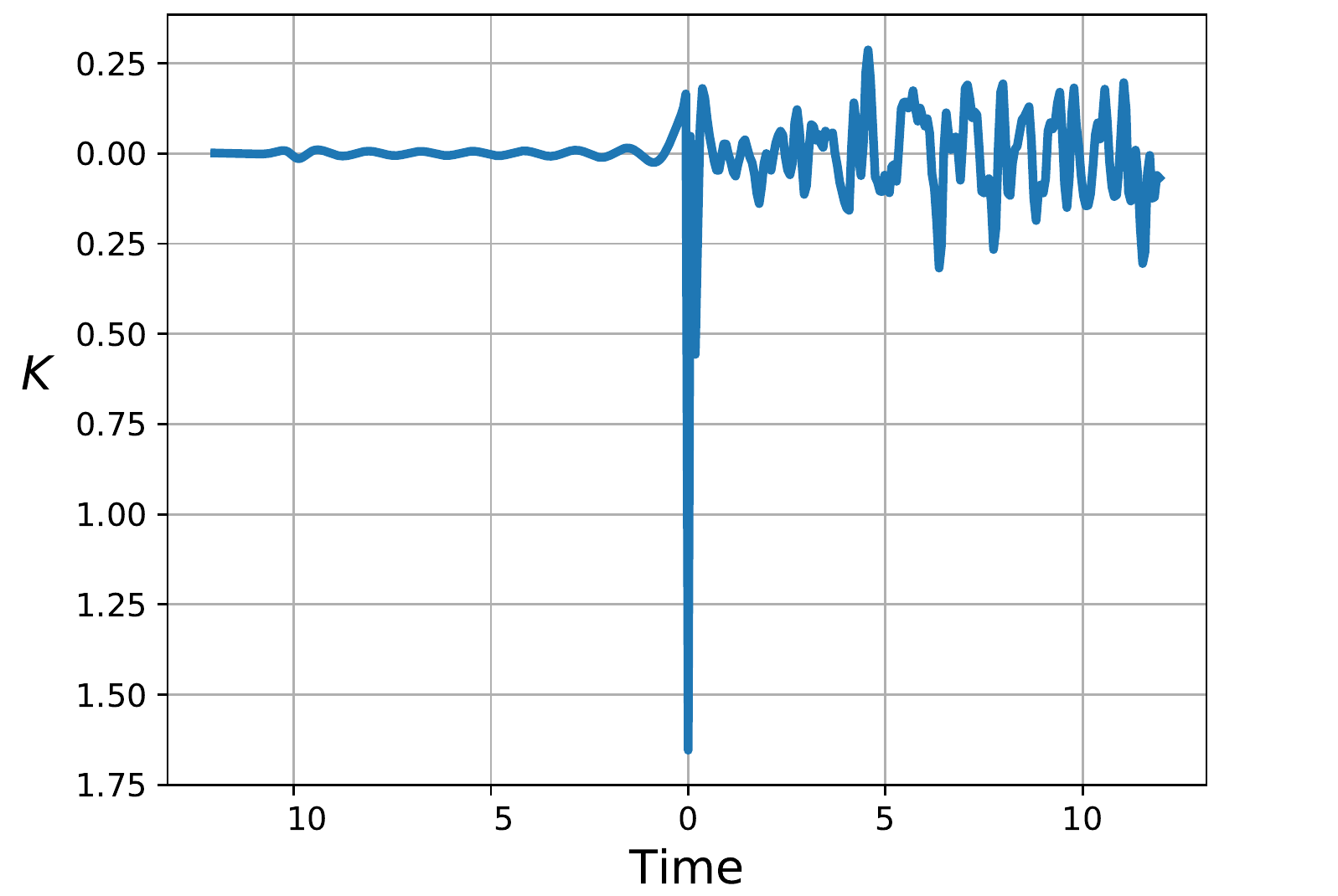}
\end{subfigure}
\caption{ (a) Surrogate response function $H_{k-2,\Psi_{1}}$. (b) Non-singular part of the surrogate response function $K_{k-2,k+1}$.}
\label{fig:more16}
\end{figure}

\begin{table}
\begin{centering}
 \begin{tabular}{c|c}
	\toprule
	$G^{(1)}$&  $G^{(1)}$ and $G^{(2)}$\\
	\hline 
    0.58 &  0.0063\\ 
    \bottomrule
\end{tabular}
\caption{PI for the surrogate response function $H_{k-2,k+1}$ when  one forcing $G^{(1)}$ is used (left) and PI for the surrogate response functions $H_{k-2,\Psi_{1}}$ and $H_{k-2,\Psi_{1}}$ when two forcings $G^{(1)}$ and $G^{(2)}$ are applied.}
\label{tab:improvement}
\end{centering}
\end{table} 

\section{Conclusions} \label{conclusions}
In this paper we have explored the possibility of using, in the linear regime, an observable of a perturbed system as predictor of the response of another observable of the same system perturbed by a forcing. While specific conditions need to be met to be able to perform an actual prediction, it is always possible to reconstruct  \textit{a posteriori} the desired signal. The procedure requires gathering first some information on the relationship between the response of the two observables . Such a knowledge can be obtained by looking at the linear response of the two observables to the forcing undergoing a broadband temporal modulation (e.g. a kick in the form of a Dirac's delta). Then, the approach allows to use the response of one observable to reconstruct and, when possible, predict, the response of the second observable for any temporal pattern of modulation of the forcing. Hence, the first observable is used as a surrogate for the external forcing. The theory clarifies that the ability of two observable to predict each other is not the same, and allows for the treatment of $N$ independent forcings to the system.

This approach can be very useful in the case we are facing an inverse problem where we have limited information on the system and in particular on the acting external forcing, while we can observe multiple features of the system at the same time and we want to be able to use internal feedbacks as surrogate forcings. This general viewpoint is closely linked to the vast class of problems associated with finding causal links in complex systems and is of potential interest in many research area dealing with complexity, such as neurosciences and geosciences. Note that the very science of paleoclimatic reconstruction and the definition of proxy data implicitly uses some of the ideas presented here. Another area of applications of surrogate response theory is  the analysis of the response of spatially extended system to perturbations, which has been the focus on this contribution. 

We have focused on the interplay between applying localised forcing and observing the system at specific locations in the vicinity of where the forcing is applied using the L96 model as benchmark system.  We have shown that, closely following the way signals propagate in the L96 model, one can establish a hierarchy of observables in terms of their ability to predict each other, where higher ranking observables are characterised by being impacted earlier by the applied perturbation.  Such a hierarchy can be analytically motivated by looking at the asymptotic properties of suitably defined response functions. The prognostic ability of an observable with respect to a predictand can be quantified by evaluating the relative weight of the causal vs. of the non-causal components of the corresponding surrogate response function. Indeed, one can also verify that, in one considers two observables that are impacted by the applied forcing with the same time delay, it is in general true that the ability to predict each other is not symmetric.

The presence of such an asymmetry is associated with the group velocity on travelling waves in the system: variables with lower index predict better variables with higher index than vice-versa. Finally, we have shown that implementing a more general form of surrogate response theory, where two independent forcings are applied and two observables are used as predictors, improves the quality of the prediction at local level even if the second observable used as predictor is a global one, because we are able to regularise the inverse problem addressed in this work compared to the simpler setting where one forcing is applied and one observable is used as  predictor. Our results are suggestive of the possibility of using the partial information gathered from the response of some observables in spatially extended system to reconstruct and predict the response for other observables when it is hard or impossible to know the exact form of the forcings.

\section*{Acknowledgements}
UMT acknowledges the support received from the EU through the ERASMUS+Traineeship program and from the Scuola Galileiana. VL acknowledges the support received from the Horizon 2020 EU project TiPES (grant No.  820970).

\appendix

\section{Predictability Index: a simple example}\label{exampleR}
\begin{figure}
\centering
\begin{subfigure}
  \centering
  \includegraphics[width=0.8\linewidth]{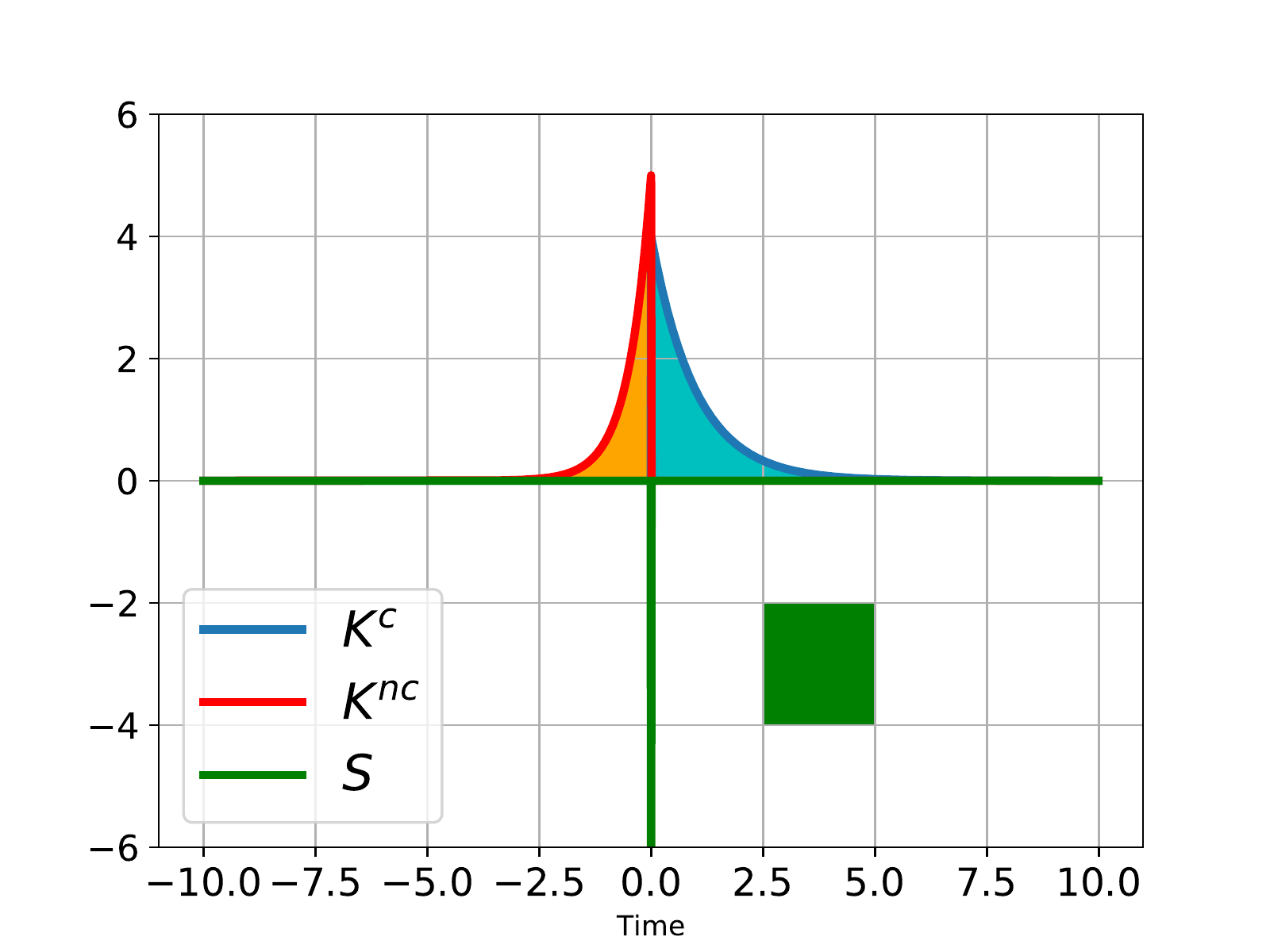}
\end{subfigure}%
\caption{Causal ($K^{c}$) and non-causal ($K^{nc}$) components of the non-singular part and singular part $S$ of the surrogate response function given in Eq. \ref{t7}. The areas under the curves correspond to the various contributions appearing in Eq. \ref{t8}. The contribution coming from the singular term is represented as a rectangle. We have used $A=4$, $B=5$, $C=-5$, $a=1$ and $b=2$; in this case, $R=5/18$.}
\label{exampleF}
\end{figure}

We now consider a simple example to show the main idea behind the effectiveness of the PI introduced in Eq. \ref{t6}. We consider the following idealised surrogate response function:
\begin{equation}
    H(t)=\Theta(t)A\,e^{-at}+\Theta(-t)B\,e^{bt}+ C\delta (t),
    \label{t7}
\end{equation}
where the coefficients $a$ and $b$ are positive while the constants $A$, $B$ and $C$ are real numbers. The non-causal and casual components of the non-singular part are respectively $K^{nc}(t)=\Theta(-t)B\,e^{bt}$ and $K^{c}(t)=\Theta(t)A\,e^{-at}$, while the singular part is $S(t)=C\delta(t)$. Note that the time scale related to the causal component is $1/a$, while the time scale related to the non-causal component is $1/b$. The smaller is the coefficient $b$, the larger is the time scale and the importance of the related component in the response function.  
The PI for the surrogate response function given in Eq. \ref{t7} is:
\begin{equation}
\begin{aligned}
 R(H)=\frac{\norm{K^{nc}(t)}_{1}}{\norm{ K^{c}(t)}_{1}+\norm{S(t)}_1}
\end{aligned}
\label{t8}
\end{equation}
where $\norm{K^{nc}(t)}_{1}=|B|/b$, $\norm{K^{c}(t)}_{1}=|A|/a$, and $\norm{S(t)}_1=|C|$.
Clearly, its value depends on the interplay of the time scales of the causal ($1/a$) and non-causal ($1/b$) components and of the magnitude of the three components $A$, $B$, and $C$; see Fig. \ref{exampleF} for a graphical representation of the contributions of the various terms. 

\section{Test of the Linearity of the Response in the Lorenz '96 model} \label{linear}

When performing an analysis based on LRT, it is clearly essential to make sure that we are indeed studying the system in a regime where higher order terms in the perturbative expansion accounting for the full response are negligible. 

\begin{figure}
\centering
\begin{subfigure}
  \centering
a)  \includegraphics[width=0.45\linewidth]{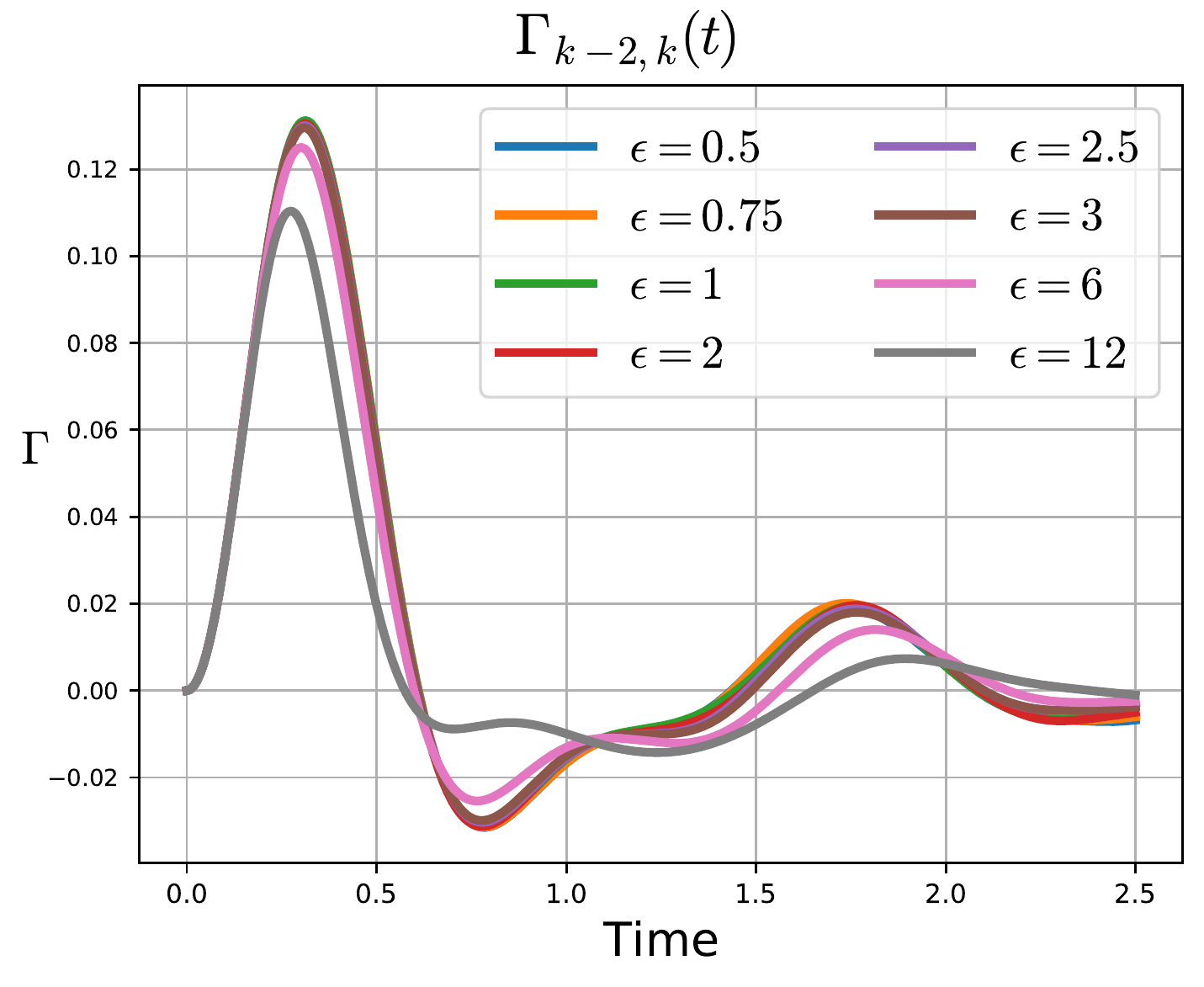}
\end{subfigure}%
\begin{subfigure}
  \centering
b) \includegraphics[width=0.45\linewidth]{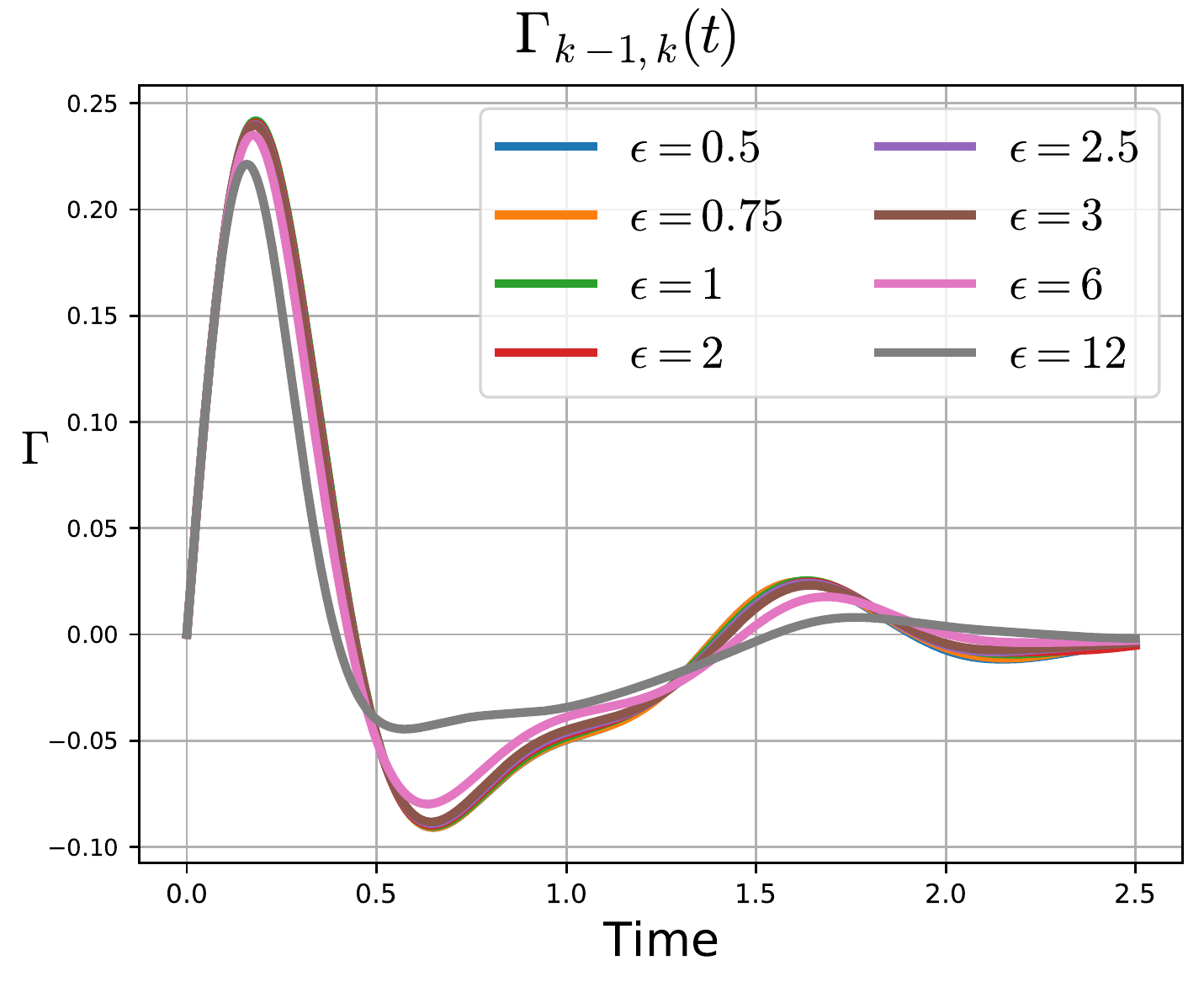}  \\
\end{subfigure}
\begin{subfigure}
  \centering
c)  \includegraphics[width=0.45\linewidth]{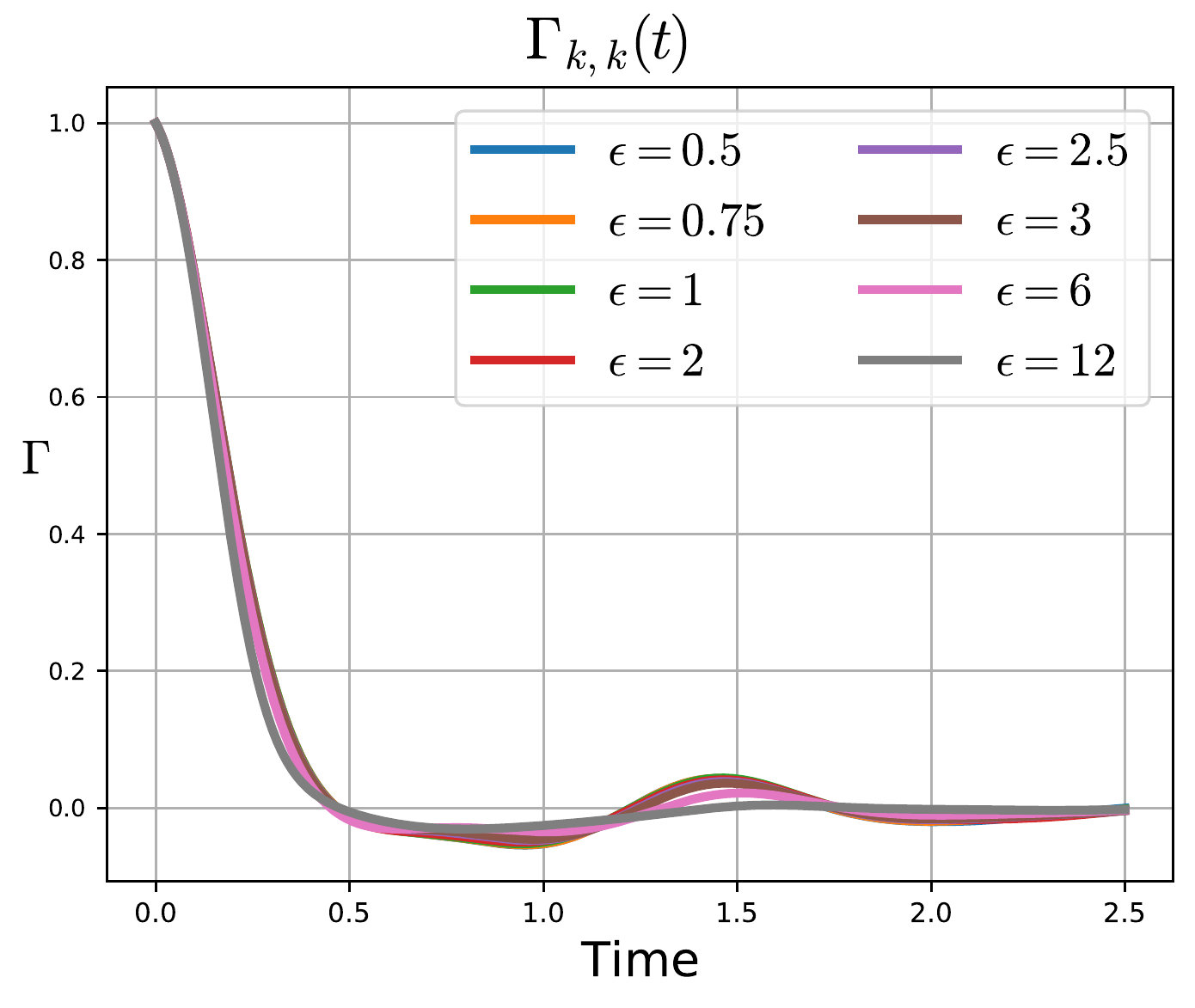}
\end{subfigure}%
\begin{subfigure}
  \centering
 d) \includegraphics[width=0.45\linewidth]{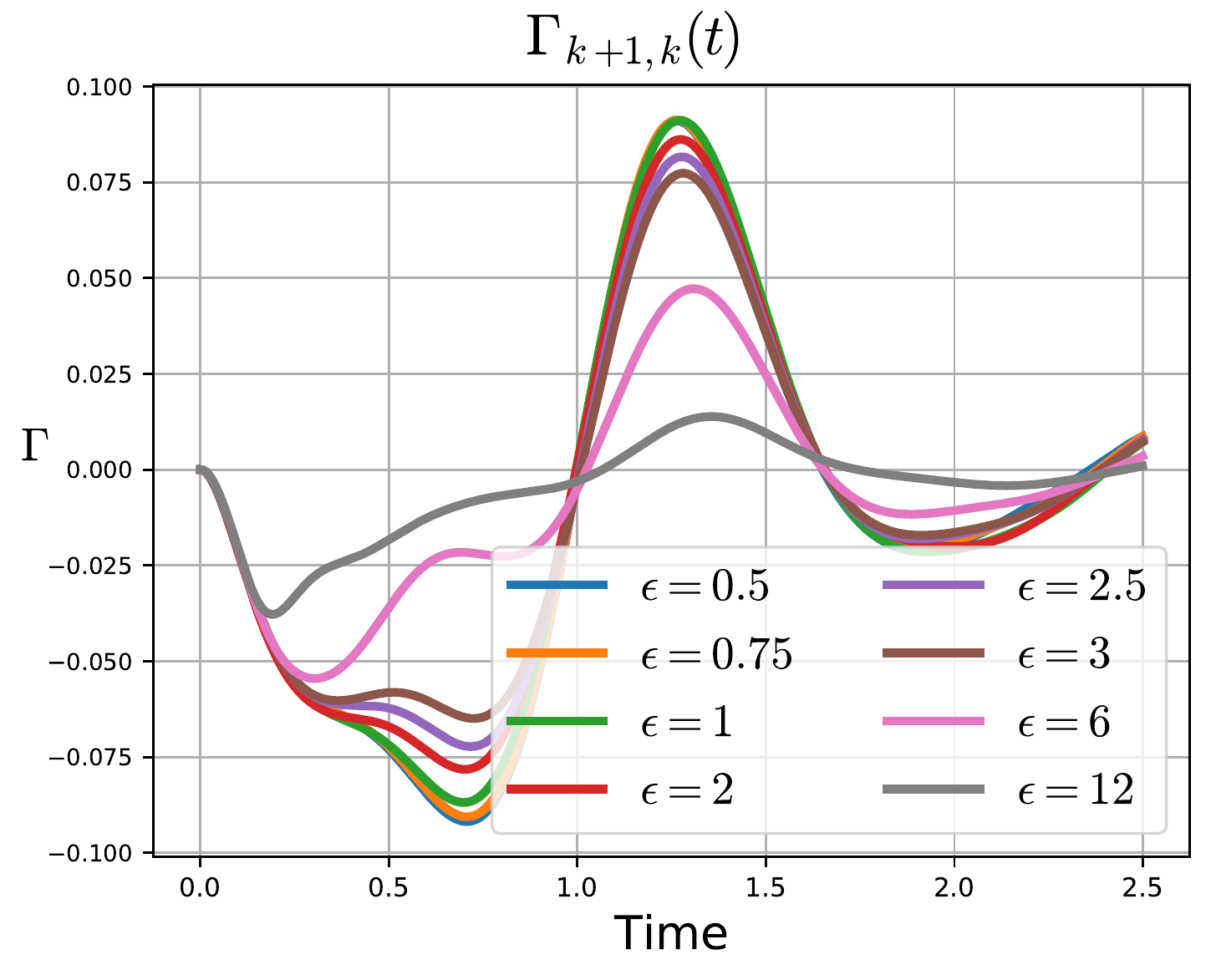} \\ 
\end{subfigure} 
  \begin{subfigure}
  \centering
e) \includegraphics[width=0.45\linewidth]{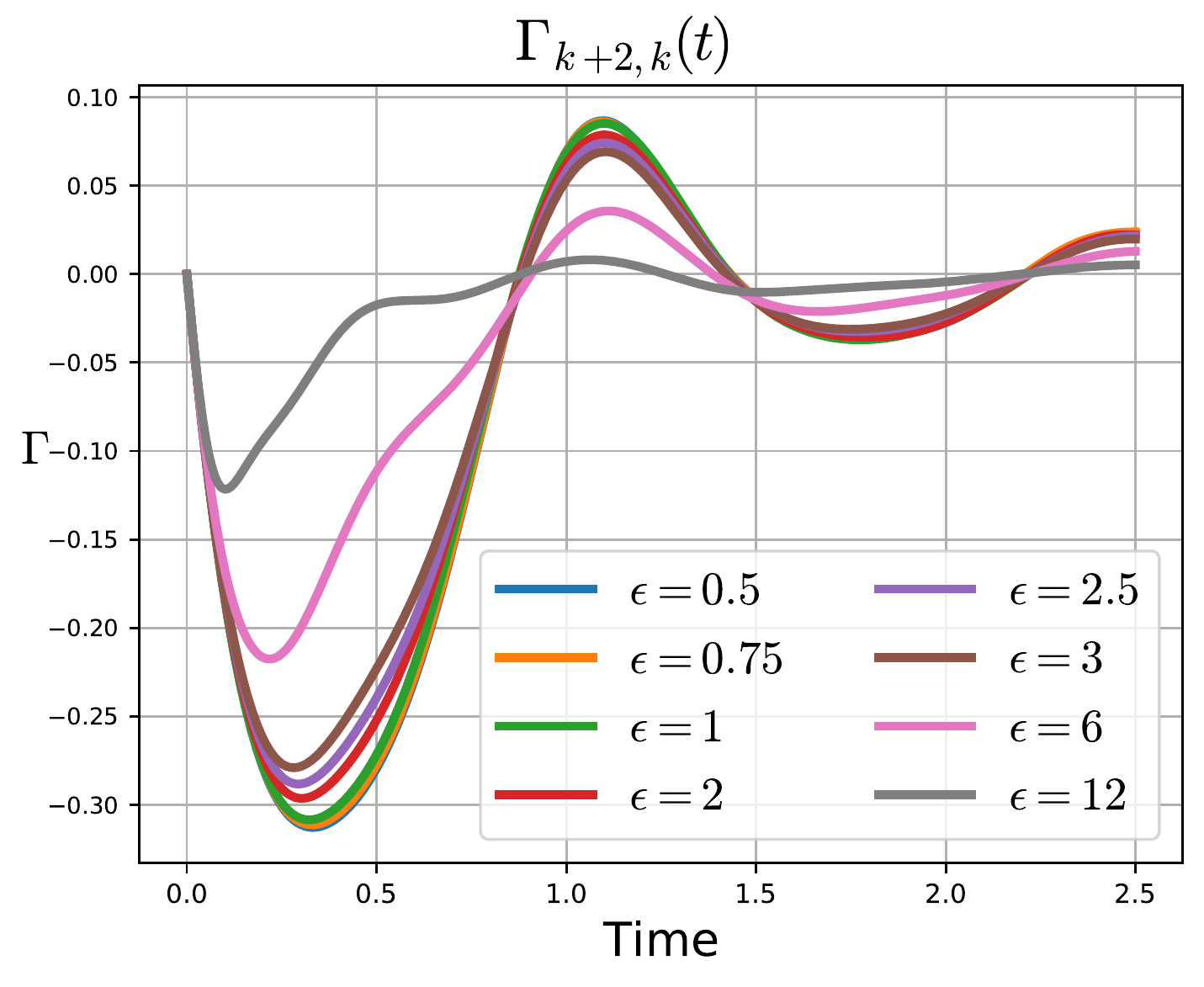}  
\end{subfigure}
\caption{Testing the linearity of the response of the system to the perturbation given in Eq. \ref{l7}. Different estimates for the response functions have been obtained by using different values of $\epsilon$. }.
\label{testlinearity}
\end{figure}

\begin{figure}
\centering
\begin{subfigure}
  \centering
a)  \includegraphics[width=0.45\linewidth]{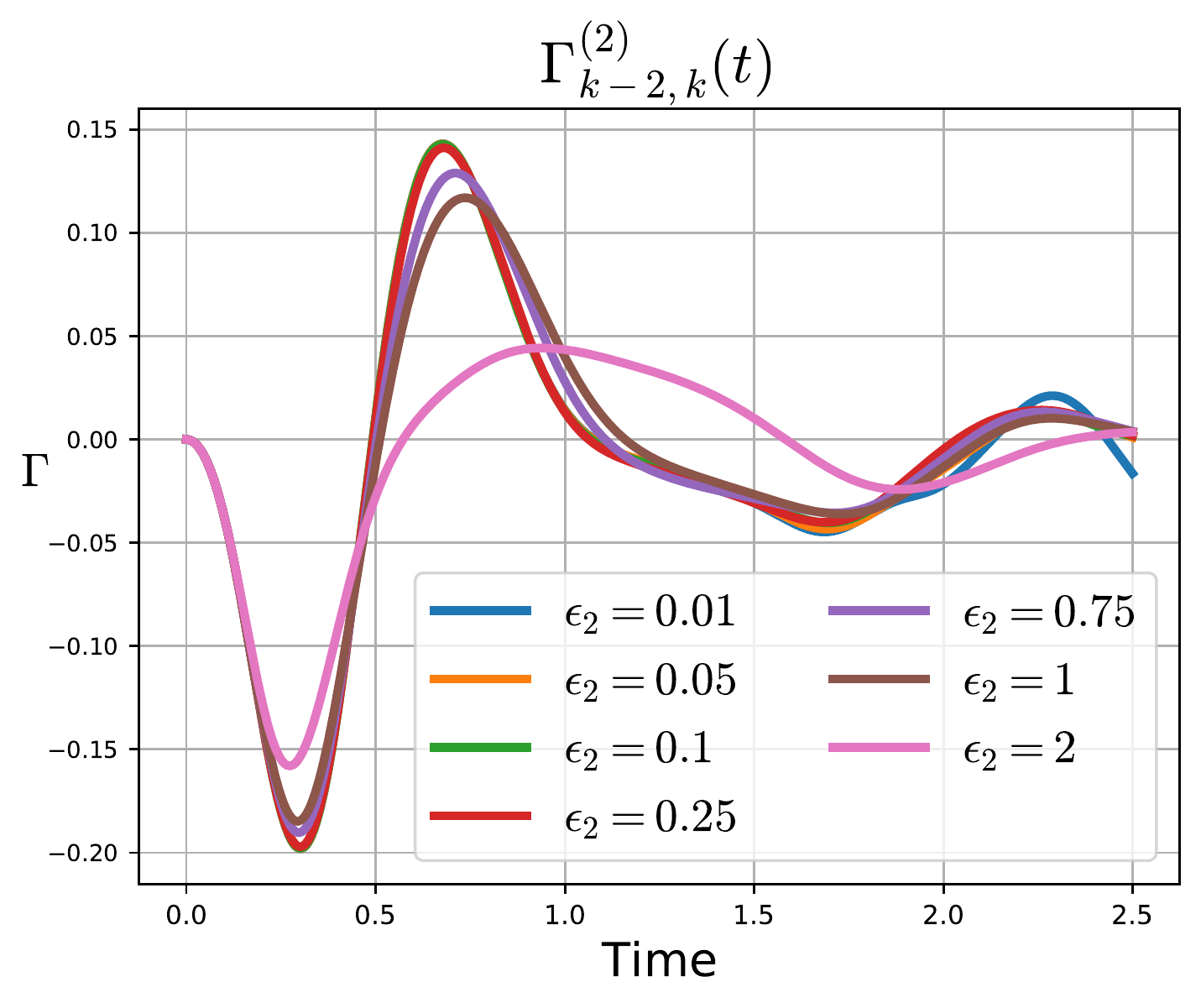}
\end{subfigure}%
\begin{subfigure}
  \centering
b) \includegraphics[width=0.45\linewidth]{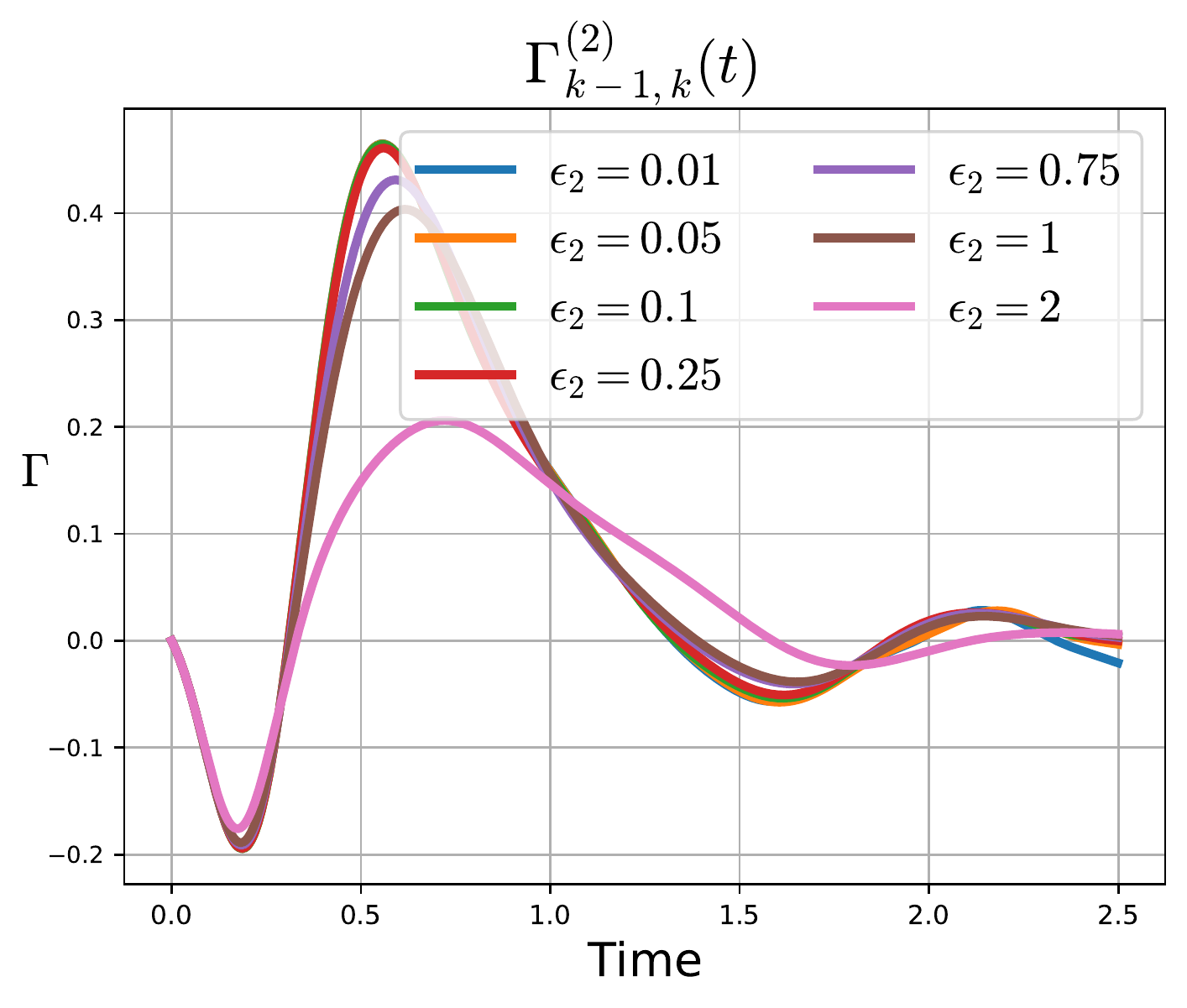} \\ 
\end{subfigure}
\begin{subfigure}
  \centering
c)  \includegraphics[width=0.45\linewidth]{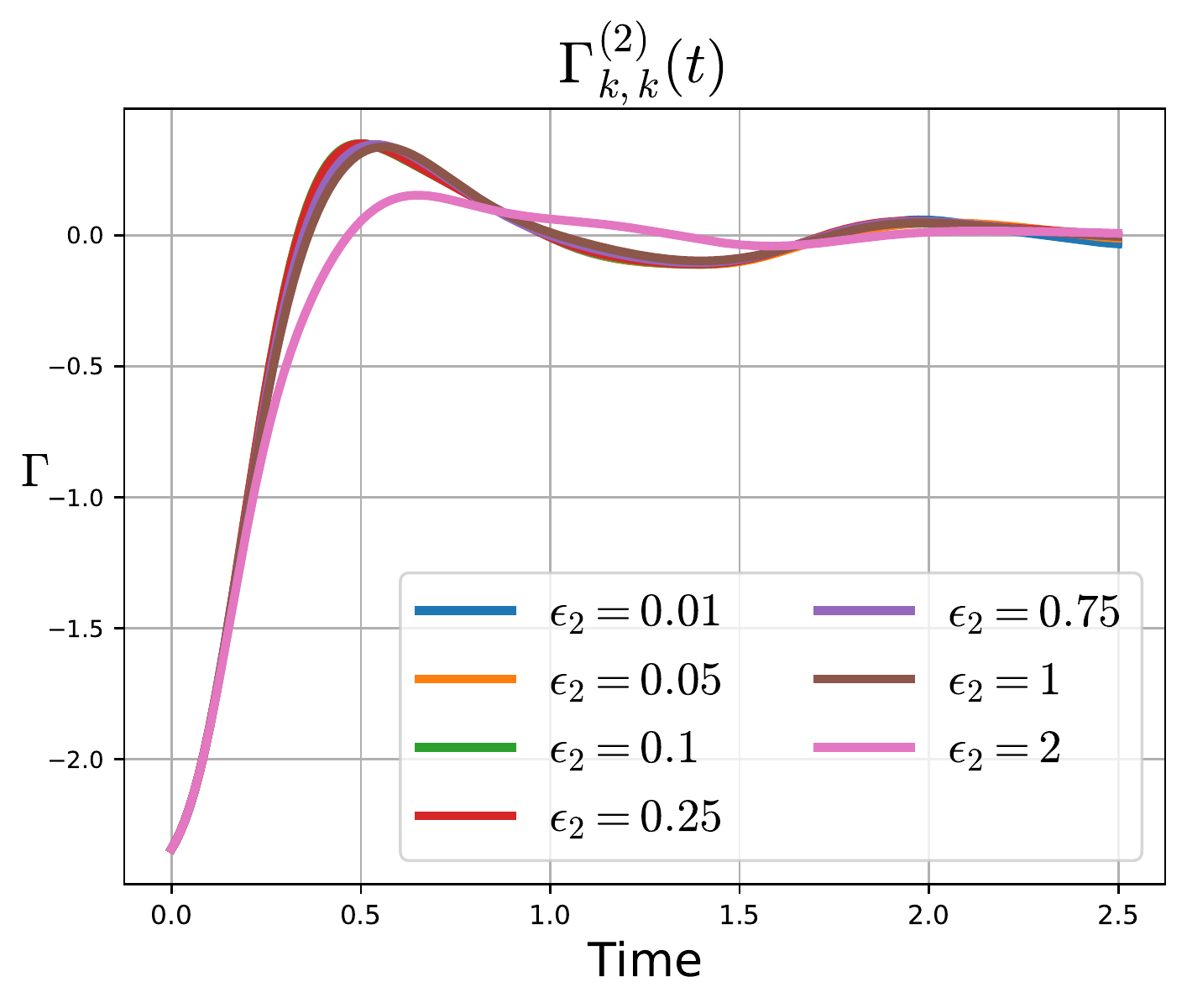}
\end{subfigure}%
\begin{subfigure}
  \centering
d) \includegraphics[width=0.45\linewidth]{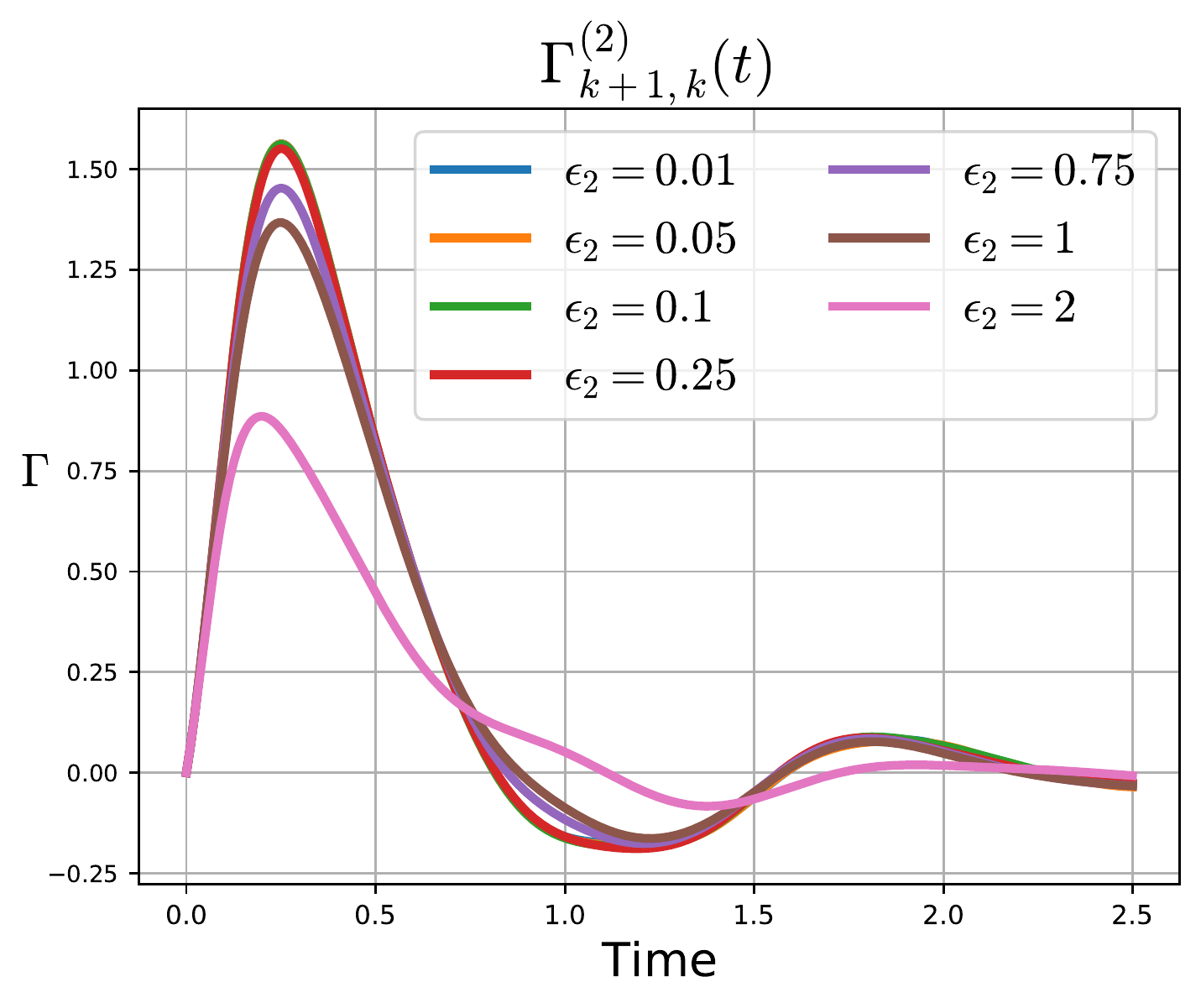}  \\
\end{subfigure} 
  \begin{subfigure}
  \centering
e) \includegraphics[width=0.45\linewidth]{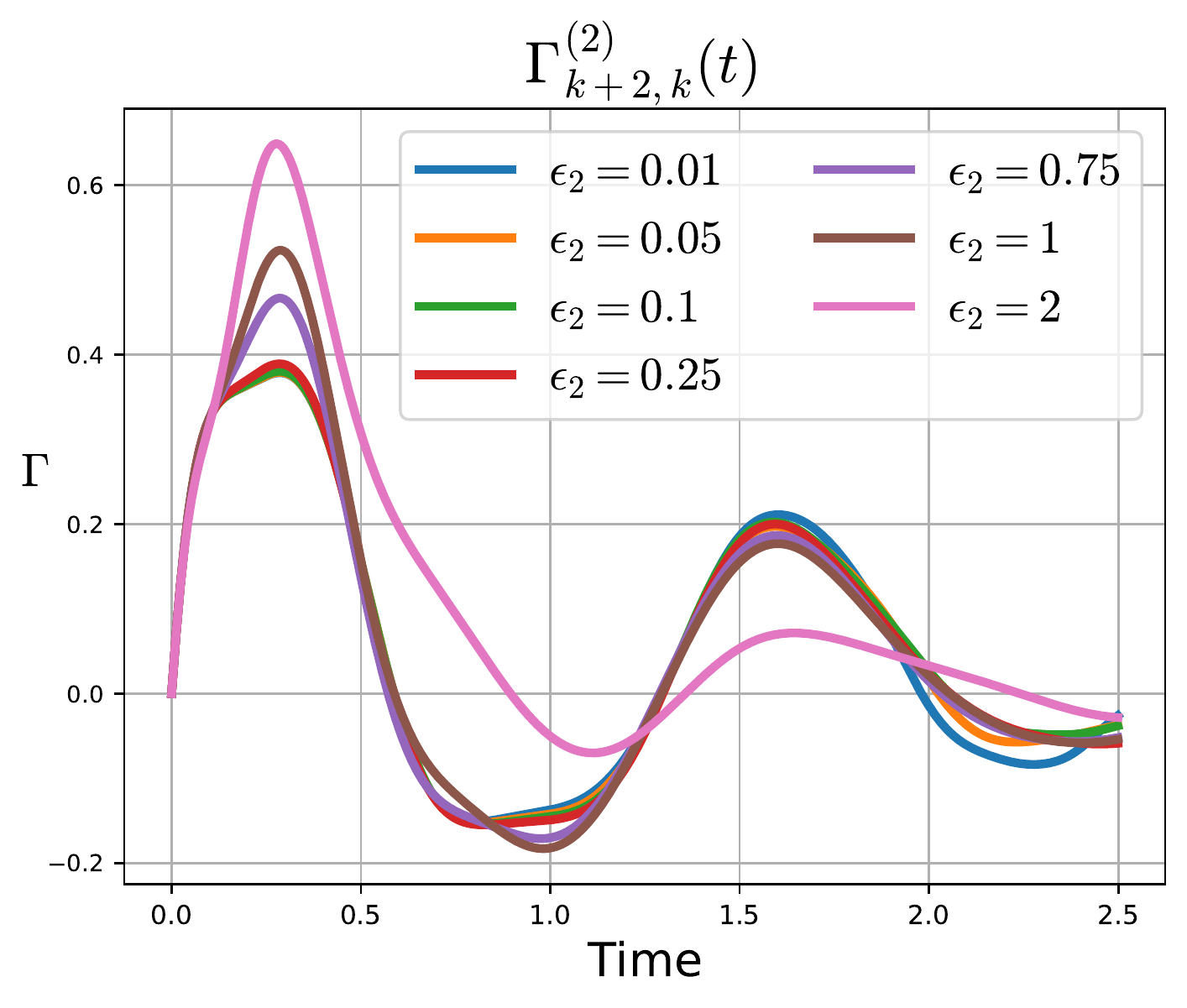}  
\end{subfigure}
\caption{Same as Fig. \ref{testlinearity}, but for the perturbation  given in Eq. \ref{n206}. }.
\label{testlinearity2}
\end{figure}

Figure \ref{testlinearity} portrays the estimates for various response functions $\Gamma_{i,j}(t)$ obtained using a range of values for the  intensity $\epsilon$ of the perturbation applied according to Eq. \ref{l7}. In all cases we have followed the procedure described in Sect. \ref{num}. The $\epsilon=1$ curves are also reported in Fig. \ref{fig:gamma_functions}. Figure \ref{testlinearity2} provides the corresponding information on the range of linearity of the response of the system to the perturbation applied according to Eq. \ref{n206}. 

\section{Asymptotic Properties of the Response in the Lorenz '96 model} \label{asymp}

It is possible to obtain the behaviour of the response functions $\Gamma_{i,k}$ for $t\rightarrow 0^{+}$ in the L96 system as follows. Let's consider the response function $\Gamma_{i,k}$ in our case, with space pattern $G(x)=\epsilon\delta_{ik}$ and perturbed observable $x_{i}$:
 \begin{equation}
 \begin{aligned}
    \Gamma_{i,k}=&\Theta(t)\int \rho_0(dx)\,\sum_{l}G\,\delta_{kl}\partial_{l}(x_{i}(t))\\
    =& \Theta(t)\int \rho_0(dx)\,\partial_{k}(x_{i}(t)),
 \end{aligned}
 \label{n0}
 \end{equation}
 where $\partial_{k}$ stands for the derivation with respect to $x_{k}(0)$ and $\rho_0$ is the steady-state distribution over the initial condition $x(0)$ from which the trajectory $x(t)$ starts. We now perform a Taylor expansion of  $x_{i}(t)$ at $t=0^{+}$ and we exploit the evolution equation Eq. \ref{l1}:
 \begin{equation}
 \partial_{k}(x_{i}(t))\approx \,\delta_{i,k}+t (C^{(1)}_{k-1}\delta_{i,k-1}+C^{(1)}_{k+1}\delta_{i,k+1}+C^{(1)}_{k+2}\delta_{i,k+2}-\delta_{ik})+\mathcal{O}(t),
 \label{exp}
 \end{equation}
 where $C^{(a)}_{p}$ is the coefficient related to the dynamical variable $p$ of the term of order $a$ in the expansion above (we have that $C^{(0)}_{k}=\delta_{ik}$).  
 The coefficients for the linear terms ($k=1$) give:
\begin{equation}
    \begin{aligned}
    C^{(1)}_{k-1}=&  x_{k-2}(0) \\
    C^{(1)}_{k+1}=&  (x_{k+2}(0)-x_{k-1}(0)) \\
    C^{(1)}_{k+2}=&   -x_{k+1}(0).
    \end{aligned}\label{cc}
\end{equation}
Considering that $\int\,\rho_0(dx)\, x_{i}= \int\,\rho_0(dx)\, x_{j}$ $\forall i,j=1,\ldots,N$ because of the discrete symmetry of the system in the unperturbed state, we have the linear term in the expansion of $\Gamma_{k+1,k}$ vanishes.

 
 We can repeat the same derivation for all the grid points of the system, taking into account that the leading term of $\Gamma_{k+2q-1,k}$ is $t^{q+1}$ instead of $t^{q}$. We obtain:
 \begin{equation}
     \Gamma_{i,k}(t)\underset{t\rightarrow 0^{+}}{\approx} \left\{
     \begin{aligned}
      \,&1 +\mathcal{O}(t), \qquad\qquad\qquad i=k \\
      \,& \langle C^{(1)}_{i}\rangle_0 t+\mathcal{O}(t^2).\,\,\,\quad i\in \{k-1,k+2\}\\
      \,&  \langle C^{(q)}_{i}\rangle_0 t^{q}+\mathcal{O}(t^{q+1}),\quad i\in\{k-q,k+2q-3,k+2q\},\, q\ge2,
     \end{aligned}
     \right.
     \label{n33}
 \end{equation}
 where the averages are over the stationary measure $\rho_0$.

\section{Singular components of the Surrogate Response}\label{sing_one}
\subsection{Case of one forcing}
We consider the perturbation with spatial pattern given by Eq. \ref{l7} and time pattern Eq. \ref{n36}. 
The $\omega\rightarrow\infty$ asymptotic behaviour of  $J_{k-1,k+2}(\omega)$ is given by:
\begin{equation}
\lim_{\omega\rightarrow \infty}J_{k-1,k+2}(\omega)= \frac{\lim\limits_{\omega \rightarrow \infty}\chi_{k-1,k}(\omega)}{\lim\limits_{\omega \rightarrow \infty}\chi_{k+2,k}(\omega)} =\frac{\langle C^{(1)}_{k-1}\rangle_0}{\langle C^{(1)}_{k+2}\rangle_0}=-1
\label{n21}
\end{equation} 
where we have used  Eqs. \ref{cc}-\ref{n33}. 
As a consequence, the non-singular component of the surrogate response function $K_{k-1,k+2}$ is obtained by performing the inverse Fourier transform of the function $J_{k-1,k+2}(\omega)+1$. 

We then consider the surrogate response functions $H_{k-2,k+1}$ and $H_{k+1,k-2}$. We proceed as above. We look at the asymptotic behaviour of the corresponding Fourier transforms:
\begin{equation}
\lim_{\omega\rightarrow\infty}J_{k-2,k+1}(\omega) = 
\frac{\lim\limits_{\omega \rightarrow \infty}\chi_{k-2,k}(\omega)}{\lim\limits_{\omega \rightarrow \infty}\chi_{k+1,k}(\omega)} =\frac{\langle C^{(2)}_{k-2}\rangle_0}{\langle C^{(2)}_{k+1}\rangle_0},
\label{n211}
\end{equation}
where $\langle C^{(2)}_{k-2}\rangle_0$ and $\langle C^{(2)}_{k+1}\rangle_0$ can be computed directly from short-time behaviour of the response functions $\Gamma_{k-2,k}$ and $\Gamma_{k+1,k}$:
\begin{equation}
    \begin{aligned}
    \langle C^{(2)}_{k-2}\rangle_0 =& \langle x_{k-3} x_{k-2} \rangle_0 \\
    \langle C^{(2)}_{k+1}\rangle_0 =& -\langle x_{k-3} x_{k-2}\rangle_0-\langle x_{k-2} x_{k}\rangle_0.
    \end{aligned}
    \label{n212}
\end{equation}
We derive the  non-singular components $K_{k-2,k+1}$ by performing the inverse Fourier transform of $J_{k-2,k+1}(\omega)-{\langle C^{(2)}_{k-2}\rangle_0}/{\langle C^{(2)}_{k+1}\rangle_0}$. Using the statistics of the unperturbed L96 model, we obtain  ${\langle C^{(2)}_{k-2}\rangle_0}/{\langle C^{(2)}_{k+1}\rangle_0}\approx-0.90$.

\subsection{Case of two forcings} 
We now want to investigate the possible presence of singular components in the surrogate response functions $H_{k-2,\Psi_{1}}$ and $H_{k-2,k+1}$, in the setting with the L96 system Eq. \ref{l1} perturbed with the two forcings introduced in Section \ref{more}. The explicit expression of their Fourier transform is given by Eq. \ref{n202} and it is the following:
\begin{equation}
   \begin{aligned}
    J_{k-2,\Psi_{1}}(\omega)=&\frac{\chi_{k+1,G^{(2)}}(\omega)\chi_{k-2,G^{(1)}}(\omega)-\chi_{k+1,G^{(1)}}(\omega)\chi_{k-2,G^{(2)}}(\omega)}{\chi_{k+1,G^{(2)}}(\omega)\chi_{\Psi_{1},G^{(1)}}(\omega)-\chi_{\Psi_{1},G^{(2)}}(\omega)\chi_{k+1,G^{(1)}}(\omega)}\\
    J_{k-2,k+1}(\omega)=&\frac{-\chi_{\Psi_{1},G^{(2)}}(\omega)\chi_{k-2,G^{(1)}}(\omega)+\chi_{\Psi_{1},G^{(1)}}(\omega)\chi_{k-2,G^{(2)}}(\omega)}{\chi_{k+1,G^{(2)}}(\omega)\chi_{\Psi_{1},G^{(1)}}(\omega)-\chi_{\Psi_{1},G^{(2)}}(\omega)\chi_{k+1,G^{(1)}}(\omega)}
    \end{aligned}
    \label{n204}
\end{equation}

To do that, we take the limit for $\omega \rightarrow \infty$ of the relations Eq. \ref{n204} and we plug the limits obtained for the response functions which appear in these relations, listed in Tab. \ref{table:limits}. As a consequence, the limit behaviour of Eq. \ref{n204} is the following:
\begin{equation}
      \begin{aligned}
    \lim\limits_{\omega \rightarrow \infty} J_{k-2,\Psi_{1}}(\omega)=& 0 \\
     \lim\limits_{\omega \rightarrow \infty}J_{k-2,k+1}(\omega)=& s_{k-2,k+1},
    \end{aligned}
    \label{n214}
\end{equation}
where: 
\begin{equation}
    s_{k-2,k+1}= \frac{-\langle x\rangle_0 \langle C^{(2)}_{k-2}\rangle_0+\langle D^{(2)}_{k-2}\rangle_0}{\langle D^{(2)}_{k+1}\rangle_0 -\langle x\rangle_0 \langle C^{(2)}_{k+1}\rangle_0 },
\end{equation}
where $\langle C^{(2)}_{k-2} \rangle_0$ and $\langle C^{(2)}_{k+1} \rangle_0 $ 
are given by Eq. \ref{n212}. Instead, 
$\langle D^{(2)}_{k-2} \rangle_0$ and $\langle D^{(2)}_{k+1} \rangle_0$ can be derived by looking at the 
$t\rightarrow 0^+$ behaviour of $\Gamma_{k+1,G^{(2)}}$ and $\Gamma_{k-2,G^{(2)}}$:
\begin{equation}
    \begin{aligned}
    \langle D^{(2)}_{k-2}\rangle_0 =& \langle x_{k-3} x_{k-2} x_{k} \rangle_0 \\
    \langle D^{(2)}_{k+1}\rangle_0 =& -3\langle x_{k-2} x_{k}\rangle_0+3\langle x_{k-1} x_{k}\rangle_0 +2\langle x_{k} x_{k+1}x_{k+3}\rangle_0-2\langle x_{k}^{2} x_{k-1}\rangle_0 -2\langle x_{k}^{2} x_{k-2}\rangle_0.
    \end{aligned}
    \label{n213}
\end{equation}
Looking at Eq. \ref{n214},  we deduce that $H_{k+1,\Psi_{1}}$ has no singular component. Instead, $K_{k-2,k+1}$ can be obtained by performing the inverse Fourier transform of  $J_{k-2,k+1}(\omega)- s_{k-2,k+1}$. Using the statistics of the unperturbed L96 model, we obtain $s_{k-2,k+1}\approx0.071$. 

\begin{table}
\begin{centering}
 \begin{tabular}{c|c}
	\toprule
	\,&  $\omega \rightarrow \infty $\\ 
	 \midrule
	$\Gamma_{\Psi_{1},G^{(1)}}$& $(iG_{1}/(N\omega)) $ \\ 
	$\Gamma_{k+1,G^{(1)}}$& $-2iG_{1}\langle C_{k+1}^{(2)}\rangle_0/\omega^{3}$  \\ 
	$\Gamma_{k-2,G^{(1)}}$& $-2iG_{1}\langle C_{k-2}^{(2)}\rangle_0/\omega^{3}$  \\ 
	$\Gamma_{\Psi_{1},G^{(2)}}$& $-(iG_{2}/(N\omega))\cdot \langle x \rangle_0$  \\ 
	$\Gamma_{k+1,G^{(2)}}$& $2iG_{2}\langle D_{k+1}^{(2)}\rangle_0/\omega^{3}$  \\ 
	$\Gamma_{k-2,G^{(2)}}$& $2iG_{2}\langle D_{k-2}^{(2)}\rangle_0/\omega^{3}$  \\
	\bottomrule
\end{tabular}
\caption{Limit behaviour for $\omega \rightarrow \infty$ of the Fourier transforms of the response functions for the observable $\Psi_{1}$, $x_{k-2}$ and $x_{k+1}$, when two forcings $G^{(1)}$ and $G^{(2)}$ are applied.}
\label{table:limits}
\end{centering}
\end{table}  

\bibliographystyle{unsrt}

\end{document}